\documentclass[12pt,preprint]{aastex}
\usepackage{graphics,graphicx,rotating,amsmath}

\newcommand{\Iobj}{{I^{obj}}}
\newcommand{\Iobs}{{I^{obs}}}
\newcommand{\IEG}{{I^{EG}}}
\newcommand{\IRCN}{{I^{RCN}}}
\newcommand{\Iobjt}{{I^{obj}(\bth)}}
\newcommand{\Iobst}{{I^{obs}(\bth)}}
\newcommand{\IRCNt}{{I^{RCN}(\bth)}}
\newcommand{\CSE}{{\Delta\theta}}
\newcommand{\bCSE}{{\Delta\bth}}
\newcommand{\DZR}{{\Delta Z_R}}
\newcommand{\WSN}{{\rm WSN}}

\newcommand{\lr}[1]{\left( #1 \right)}
\newcommand{\lrs}[1]{\left[ #1 \right]}
\newcommand{\lrabs}[1]{\left| #1 \right|}
\newcommand{\Biglr}[1]{\Bigl( #1 \Bigr)}
\newcommand{\Bigglr}[1]{\Biggl( #1 \Biggr)}
\newcommand{\Bigglrs}[1]{\Biggl[ #1 \Biggr]}
\newcommand{\Real}[1]{{\rm Re}\left[ #1 \right]}
\newcommand{\Img}[1]{{\rm Im}\left[ #1 \right]}

\newcommand{\cH}{{\cal H}}
\newcommand{\bde}{{\boldsymbol \delta}}
\newcommand{\bth}{{\boldsymbol \theta}}
\newcommand{\cC}{{\cal C}}
\newcommand{\tcC}{{\tilde{\cal C}}}
\newcommand{\cCX}{{\cal C}X}
\newcommand{\DDT}{{\Theta}}
\begin{document}
\title{Elliptical Weighted HOLICs for Weak Lensing Shear Measurement\\
part3:Random Count Noise Effect for Image's Moments in Weak Lensing Analysis}
\author{Yuki Okura\altaffilmark{1}} 
\email{yuki.okura@nao.ac.jp}

\author{Toshifumi Futamase\altaffilmark{2}}
\email{tof@astr.tohoku.ac.jp}

\altaffiltext{1}
 {National Astronomical Observatory of Japan, Tokyo 181-8588, Japan}
\altaffiltext{2}
 {Astronomical Institute, Tohoku University, Sendai 980-8578, Japan}

\begin{abstract}
This is the third paper on the improvements of systematic errors in our 
weak lensing analysis using an elliptical weight function, called E-HOLICs. 
In the previous papers we have succeeded in avoiding error which depends on ellipticity of background image.
In this paper, we investigate the systematic error which depends on signal to noise ratio of background image.
We find that the origin of the error is the random count noise which comes from Poisson noise of sky counts.
Random count noise makes additional moments and centroid shift error,
and those 1st orders are canceled in averaging, but 2nd orders are not canceled. 
We derived the equations which corrects these effects in measuring moments and ellipticity of the image 
and test their validity using simulation image. 
We find that the systematic error becomes less than 1\% in the measured ellipticity for objects 
with $S/N>3$. 
\end{abstract}

\section{Introduction}
The importance of the weak lensing analysis is now widely recognized because it has a potential to provides 
us a direct and unbiased information on the mass distribution for lens objects.   
The weak lensing analysis measures shapes(called ellipticity which has two components interpreted as direction and magnitude) of many background images(galaxies) and then averaged over an appropriate number of images to get rid of intrinsic random ellipticity of images and to withdraw the ellipticity due to gravitational tidal effect(shear) of the lensing object. The shear carries the information of mass structure of the lensing object. Thus an accurate shape measurement of the background images is critically important to accurately measure the mass distribution. 
So far weak lensing is very successful for cluster lensing (ellipticity due to shear is of 
the order of 5\%) and provides us a rich information of mass structures of clusters and of our understanding structure formation in the universe. 
Recently the cosmic shear, i.e. the weak lensing due to large scale structure(LSS) attracted much attention because of it's ability to study the nature of dark energy which is supposed to be the source of the accelerated expansion of the universe. In fact several projects for the cosmic shear measurement are proposed and some of them is almost ready to start the observation
(Hyper Suprime-Cam http://www.naoj.org/Projects/HSC/HSCProject.html, Dark Energy Survey http://www.darkenergysurvey.org/, Euclid http://sci.esa.int/euclid and so on). 
However the signal of cosmic shear is very weak(of the order of 1\%) compared with cluster lensing and thus needs special treatment.  Namely we needs to develop very accurate shape measurement scheme which avoids various systematic errors. For example the measured gravitational shear depends on the ellipticity and signal to noise of background image. Usually such dependence becomes small by averaging many of the images, but it is critically important to realize that these dependence somehow correlated with the redshift distribution of image which is also important to have an accurate measurement of the shear. Thus we cannot make a simple averaging over the images without having a method free from such systematic errors.  
The required accuracy for the measurement of ellipticity is less than 1\% in order 
to have an useful information of dark energy. 

There have been many studies in this direction and various measurement schemes are proposed(Kaiser et al 1995, Bernstein \& Jarvis 2002; Refregier 2003; Kuijken et al. 2006; Miller et al. 2007; Kitching et al. 2008; Melchior 2011). 
The accuracy of these methods are tested using the simulation data provided by STEP1(Heymans et al 2006), STEP2(Massey et al 2007), GREAT08(Bridle et al 2010) and GREAT10(Kitching et al 2012).
Although much progress is reported, none of the method achieved the required accuracy and are free from various 
systematic errors.

We have also developed a new scheme based on KSB method(Kaiser et al 1995) using an elliptical window function(we called E-HOLICs) to measure the background image as accurately as possible(Okura and Futamase 2011, Part I paper).  
It is shown in our Part II papers that the E-HOLICs can improve the systematic error which depends on ellipticity. In this paper we study the systematic error which depends on signal to noise ratio(SN).
There are some studies about this systematic error
(Hirata et al 2004, Kacprzak et al 2012, Refregier et al 2012, Okura and Futamase 2012 Part II paper, Melchior and Viola 2012).
these results show this systematic error comes from random count noise(RCN). 
Because, 1st order effects from RCN are canceled by averaging, but 2nd order effects are not canceled.
We calculate the 2nd order effects to obtain the correction formulas in
the measurement of moments and ellipticity  for Gaussian weighted images
in KSB method and E-HOLICs method (i.e. without PSF correction). Using
the simulation data GREAT08, we find that the derived formula
correct the SN dependent bias within 1\% for images with SN $\geq$ 3.  

The paper organized as follows.
In section 2, we explain and define our notations and some of the definitions.
In section 3, we calculate the 2nd order effects of RCN and obtain general formulas to correct the effects. 
We test the formula in the case of KSB method with Gaussian weight function.
The correction formula in the case of E-HOLICs is presented in section 4, and tested it using GREAT 08 simulation and find that the systematic error becomes less than 1\% in the measured ellipticity for objects 
with $S/N>3$.
In section 5, we summaries our results.
\section{Basis and Definitions}
In this section, we present notations and definitions we use in E-HOLICs method.
Some of them were defined in part2, but we add the effect of "random count noise"(hereafter RCN) and "centroid shift error"(hereafter CSE).

\subsection{Random Count Noise}
First, we write the observed brightness distribution of object as "$\Iobst$",
which is the sum of object "$\Iobjt$" and RCN "$\IRCNt$", so
\begin{eqnarray}
\Iobst=\Iobjt+\IRCNt,
\end{eqnarray}
where, "$\bth$" is position angle in complex coordinate whose origin is at the centroid of object "$\Iobjt$"
\begin{eqnarray}
\bth&\equiv&\theta^1_1\equiv\theta_1+i\theta_2,
\end{eqnarray}
and the products of the positions are notated as 
\begin{eqnarray}
\theta^N_M&=&\lr{\theta^1_1}^{\frac{N+M}{2}}\lr{\theta^{1*}_1}^{\frac{N-M}{2}},
\end{eqnarray}
N means order and M means spin-number.

We assume that RCN is Poisson noise of sky counts and also assume that all pixels have same root-mean-square(RMS) of RCN "$\sigma_{RCN}$".
We don't consider Poisson noise of objects itself in this paper. 
If an object has a photon count $N_{obj}$, then the Poisson noise is of the order of $\sqrt{N_{obj}}$,
and thus the order of errors reduces by $1/\sqrt{N_{obj}}$.
Therefore if the object is bright enough to be able to neglect sky noise,
we can also neglect own Poisson noise. 
On the other hand, if an object is faint, its Poisson noise is much smaller than the Poisson noise from sky $N_{sky}$($N_{obj}\ll N_{sky}$),
so we can neglect it.
However it needs another consideration for the situation with $N_{obj}\sim N_{sky}$ which will be discussed 
in other paper.

\subsection{Notations}
In measuring moments of image by E-HOLICs method,
we use an elliptical Gaussian weight function with ellipticity "$\bde_W \equiv \delta_{W1} + i\delta_{W2}$"
for measuring the complex moments, and we define this weight function as
\begin{eqnarray}
W(\bth,\bde_W)&\equiv&e^{-\frac{\theta^2_0-\Real{\bde_W^*\theta^2_2}}{\sigma_W^2}},
\end{eqnarray}
where $\sigma_W^2$ is a size parameter of weight function.

The complex moments and HOLICs of arbitrary brightness distribution without centroid shift error(CSE) 
are defined as
\begin{eqnarray}
\label{eq:defCM}
Z^N_M(I,\bde_W)&\equiv&\int d^2\theta \theta^N_M I(\bth)W(\bth,\bde_W)\\
\cH^N_M(I,Z^O_P,\bde_W)&\equiv&\frac{Z^N_M(I,\bde_W)}{Z^O_P(I,\bde_W)}.
\end{eqnarray}

In this paper, we define the origin of the coordinate at the centroid of $\Iobjt$, 
therefore 
\begin{eqnarray}
\label{eq:Z11obj}
Z^1_1(\Iobj,\bde_W)&\equiv&0.
\end{eqnarray}
However, RCN causes CSE,
so we cannot obtain eq.(\ref{eq:Z11obj}) in real analysis.
We notate CSE due to RCN as $\bCSE = \CSE^1_1$.
then the complex moments with RCN and CSE that we measure in real analysis are defined as 
\begin{eqnarray}
\label{eq:DeltaCM}
\hat Z^N_M(\Iobs,\bde_W)&\equiv&\int d^2\theta \lr{\theta-\CSE}^N_M \lr{\Iobjt+\IRCNt}W(\bth-\bCSE,\bde_W),
\end{eqnarray}
and HOLICs are measured as
\begin{eqnarray}
\label{eq:DeltaHOLICs}
\hat \cH^N_M(\Iobs,Z^O_P,\bde_W)&\equiv&\frac{\hat Z^N_M(\Iobs,\bde_W)}{\hat Z^O_P(\Iobs,\bde_W)}.
\end{eqnarray}
The detail of this CSE is expressed in section 3.

\subsection{WSN}
Here, we define weighted signal to noise ratio "WSN"
with elliptical weight function as
\begin{eqnarray}
\WSN\equiv\frac{\int d^2\theta \Iobjt W(\bth,\bde_W)}{\sigma_{RCN}\sqrt{\int d^2\theta W(\bth,\bde_W)}}
=\frac{Z^0_0(\Iobj,\bde_W)}{\sigma_{RCN}\sqrt{S_W}}
\approx\frac{Z^0_0(\Iobs,\bde_W)}{\sigma_{RCN}\sqrt{S_W}},
\end{eqnarray}
where $S_W$ is an integral of weight function or weighted area
\begin{eqnarray}
S_W&=&\frac{\sigma^2_W\pi}{\sqrt{1-\delta_W^2}}\\
\delta_W&\equiv&|\bde_W|.
\end{eqnarray}
WSN appears frequently in the following calculations, 
so we use WSN instead of SN.

\begin{figure*}[htbp]	
\epsscale{1.0}
\plotone{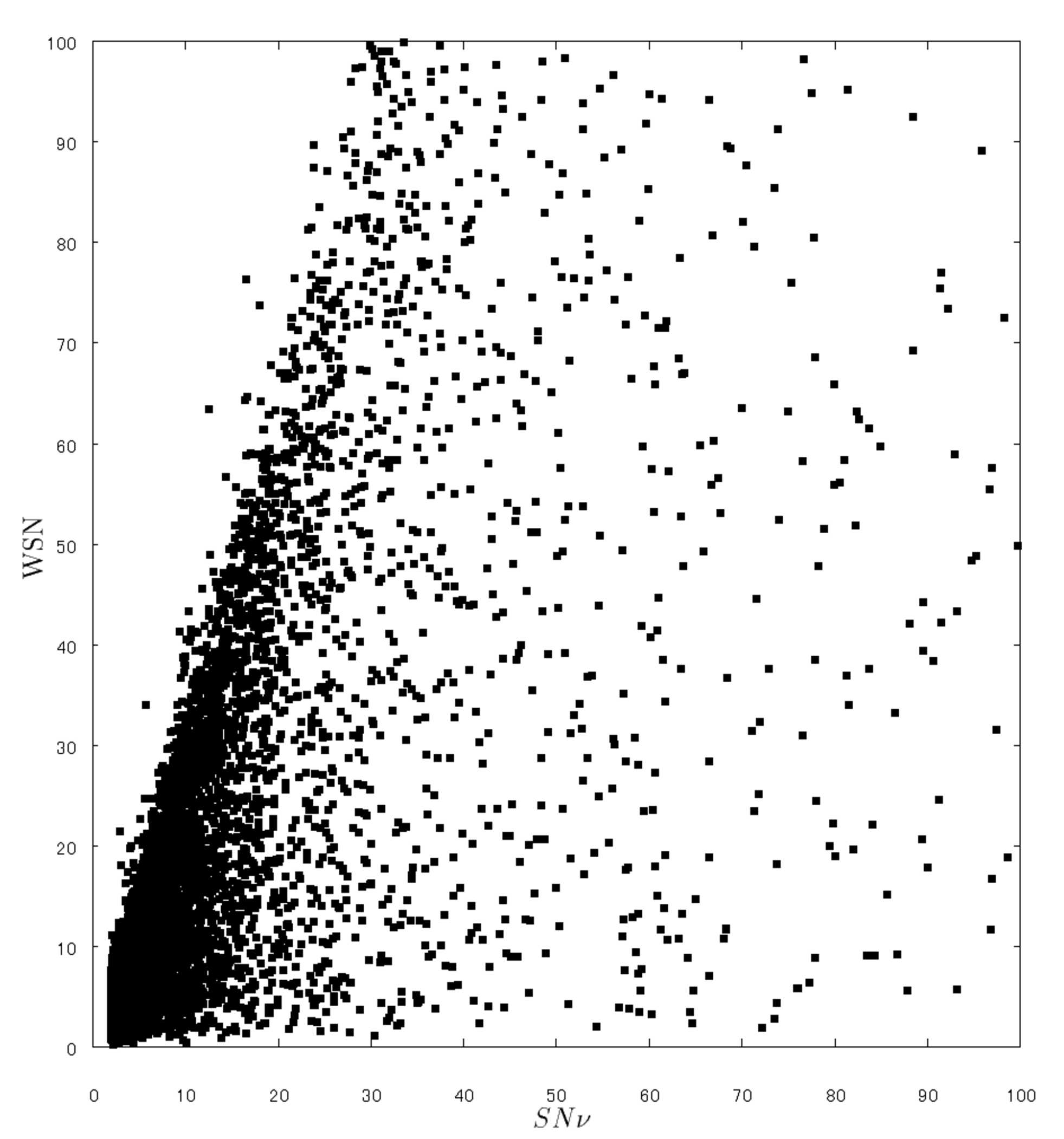}
\caption{
\label{fig:WSN}
Plots of SN$_\nu$ .v.s. WSN. We used background objects of Abell 1689 real data.
We reject objects which have SN lower than 2.
We can see some objects have correlation about $WSN= 3SN_\nu$ and almost objects have $WSN\leqq 3SN_\nu$.
} 
\end{figure*}

\begin{figure*}[htbp]	
\epsscale{1.0}
\plotone{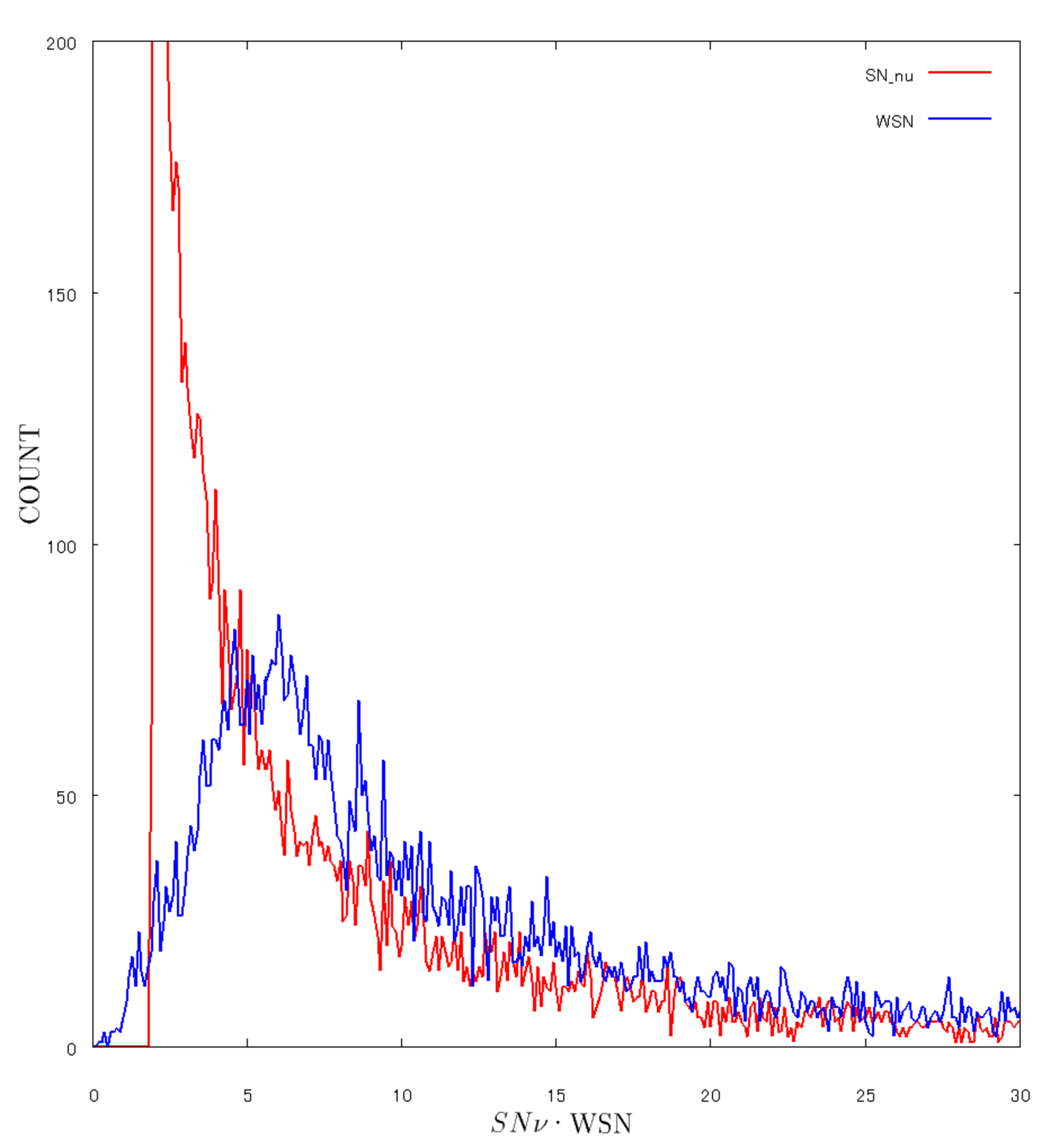}
\caption{
\label{fig:WSN_DISTRIBUTION}
Distributions of counts of SN and WSN. We used background objects of Abell 1689 real data.
We reject objects which have SN lower than 2.
} 
\end{figure*}

We measure SN and WSN of back ground objects 
detected from Abell 1689 real data taken by Subaru suprime-cam,
and we use only objects having $\nu\ge7$ by IMCAT(http://www.ifa.hawaii.edu/~Kaiser/imcat) detection.
And we define a signal to noise ratio "SN$_\nu$" defined from $\nu$ as
\begin{eqnarray}
SN_\nu\equiv\frac{\nu}{3.5}.
\end{eqnarray}
The plots of $SN\nu$ and WSN are shown in fig.\ref{fig:WSN} and
we can see the following relation. 
\begin{eqnarray}
WSN\leqq 3SN_\nu.
\end{eqnarray}
Fig.\ref{fig:WSN_DISTRIBUTION} shows the count distributions of $SN_{\nu}$ and WSN.

\subsection{Averaging}
In weak lensing analysis, usually "averaging" means the averaging over a parameter of several different objects,
but in this paper "averaging" means averaging over a parameter of same object but different RCN.
It means that we observe the same object many times.
Therefore the difference between the value measured without RCN and the averaged value with RCN is the  systematic error,

Because we use different RCN, the 
averaged complex moments of them vanish,
\begin{eqnarray}
\overline{Z^N_M(\IRCN,\bde_W)}&=&0.
\end{eqnarray}
So averaged values of complex moments of RCN are 0,
but squares of the moments are not 0,
because RCN has self-correlation.

Standard deviation of RCN at arbitrary position $\bth_a$ is obtained as
\begin{eqnarray}
\frac1{N}\sum^N_i I^{RCN}_i(\bth_a)I^{RCN}_i(\bth'_a) = \overline{I^{RCN}(\bth_a)I^{RCN}(\bth'_a)} = \sigma_{RCN}^2\delta_D\lr{\bth_a-\bth'_a},
\end{eqnarray}
where "i" means $i$th set of RCN and $\delta_D(\bth)$ is Dirac delta Function..
Let $G^N_M$ be defined as a square of complex moments of $\IRCNt$ as 
\begin{eqnarray}
G^{N+O}_{M+P}(\bde_W)\equiv\frac{Z^N_M(\IRCN,\bde_W)Z^O_P(\IRCN,\bde_W)}{\sigma_{RCN}^2},
\end{eqnarray}
and average value of $G^N_M$ is obtained as
\begin{eqnarray}
\overline{G^{N+O}_{M+P}(\bde_W)}
&=&
\frac{1}{\sigma_{RCN}^2}\overline{Z^N_M(\IRCNt,\bde_W)Z^O_P(\IRCNt,\bde_W)}
\nonumber\\&=&
\frac{1}{\sigma_{RCN}^2}\overline{\int d^2\theta \theta^N_M \IRCNt W(\bth,\bde_W)\int d^2\theta' \theta'^O_P I^{RCN}(\bth') W(\bth',\bde_W)}
\nonumber\\&=&
\frac{1}{\sigma_{RCN}^2}\int d^2\theta \theta^N_M W(\bth,\bde_W)\int d^2\theta' \theta'^O_P W(\bth',\bde_W) \delta_D\lr{\bth-\bth'}
\nonumber\\&=&
\int d^2\theta \theta^{N+O}_{M+P} \lr{W(\bth,\bde_W)}^2.
\end{eqnarray}
Because we use elliptical Gaussian for weight function $W(\bth,\bde_W)$,
$\overline{G^N_M(\bde_W)}$ can be calculated analytically.
The detailed values of $\overline{G^N_M(\bde_W)}$ can be seen in Appendix \ref{AP:GNM}.
Then we can obtain the averaged value of the square of the complex moments as the 
product of $\sigma^2_{RCN}$ and $\overline{G^N_M(\bde_W)}$.

\begin{figure*}[htbp]	
\epsscale{1.0}
\plotone{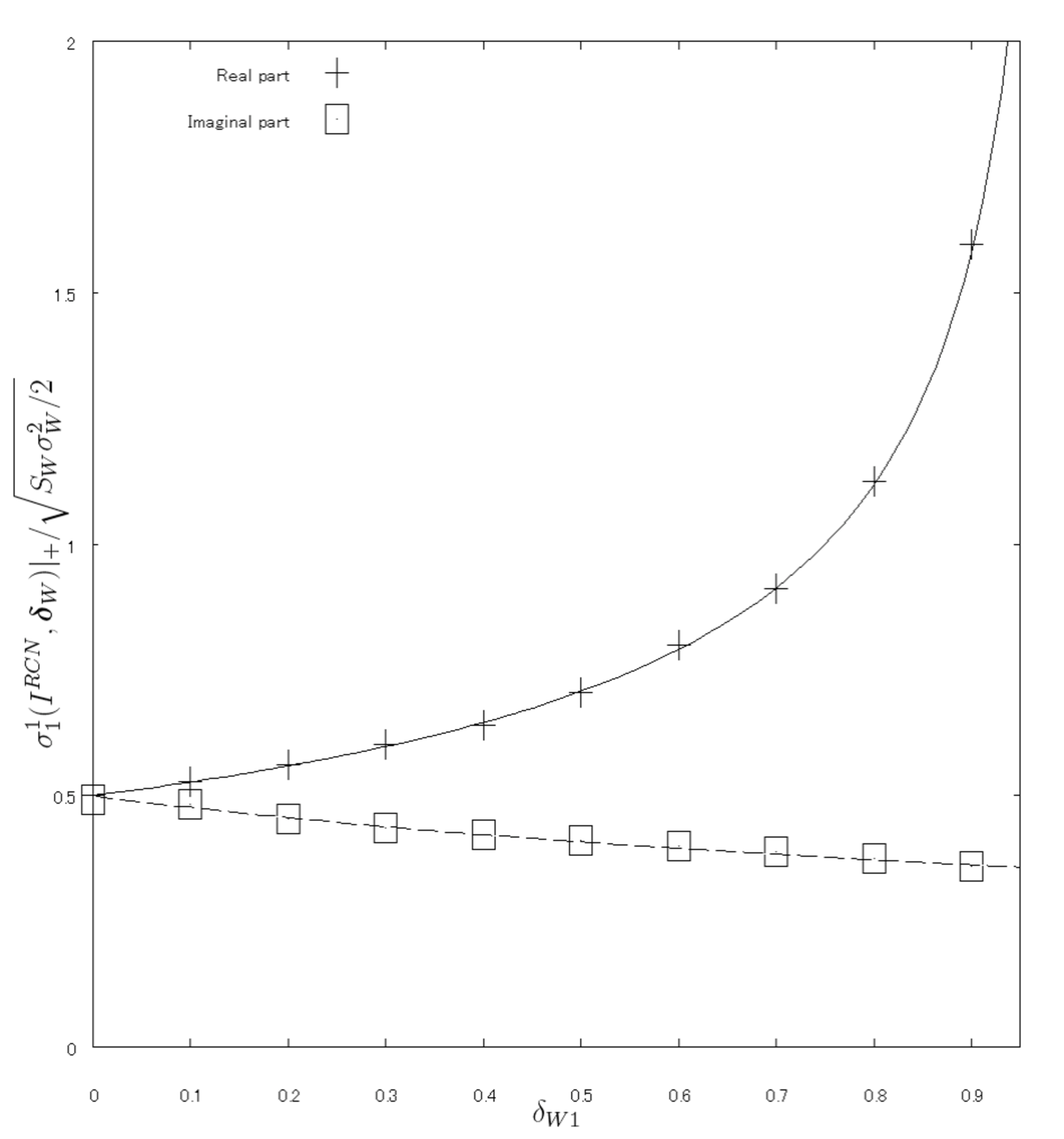}
\caption{
\label{fig:E-HOLICs_3_sigma11}
Test of standard deviations of $Z^1_1(\IRCN,\bde_W)$ with normalization by $\sqrt{S_W\sigma_W^2/2}$ and $\sigma_{RCN}=1$.
Horizontal axis means ellipticity of weight function $\delta_{W1}$ ($\delta_{W2}=0$) and vertical axis means a standard deviation of the complex moments.
Cross(square) points mean real(imaginal) part of $\sigma^1_1$ 
and line(dash) means real(imaginal) part of analytical estimation eq.(\ref{eq:RCNRMS11})
} 
\end{figure*}

Here we show explicitly the calculation for $\sigma^1_1(I^{RCN},\bde_W)|_+$ which is the standard deviation of each components of $Z^1_1(\IRCN,\bde_W)$,.
\begin{eqnarray}
\label{eq:RCNRMS11}
\sigma^1_1(I^{RCN},\bde_W)|_+&\equiv&\sqrt{\overline{\lr{\Real{Z^1_1(I^{RCN},\bde_W)}}^2}}+i\sqrt{\overline{\lr{\Img{Z^1_1(I^{RCN},\bde_W)}}^2}}
\nonumber\\&=&
\sqrt{\sigma_{RCN}^2\int d^2\theta \lr{\theta_1 W(\bth,\bde_W)}^2}+i\sqrt{\sigma_{RCN}^2\int d^2\theta \lr{\theta_2 W(\bth,\bde_W)}^2}
\nonumber\\&=&
\sigma_{RCN}\lr{\sqrt{\frac{\overline{G^2_0(\bde_W)}+\Real{\overline{G^2_2(\bde_W)}}}{2}}+i\sqrt{\frac{\overline{G^2_0(\bde_W)}-\Real{\overline{G^2_2(\bde_W)}}}{2}}}
\nonumber\\&=&
\sigma_{RCN}\frac{\sqrt{S_W\sigma_W^2}}{2\sqrt{1-\delta_W^2}}\lr{\sqrt{\frac{1+\delta_{W1}}{2}}+i\sqrt{\frac{1-\delta_{W1}}{2}}},
\end{eqnarray}
fig.\ref{fig:E-HOLICs_3_sigma11} shows simulation results of eq.(\ref{eq:RCNRMS11}) with normalization by $\sqrt{S_W\sigma_W^2/2}$ and $\sigma_{RCN}=1$.

\section{Centroid Shift Error and Complex Moments with Random Count Noise}
In this section, we present calculations of centroid shift error(CSE) and the complex moments with random count noise(RCN) in detail,
where we assume the brightness distribution of object as an elliptical Gaussian image and adopt 
KSB method.

\subsection{Centroid Shift Error with Random Count Noise}
We present calculations about CSE.
In this paper, we define true centroid as the origin of the complex coordinate, so $\bth=0$ is true centroid,
however centroid of $\Iobst$  we measure is different from the origin and this difference is CSE "$\Delta\bth$".

We measure the centroid as a position which vanishes the dipole moment of $\Iobst$, so the dipole moments of $\Iobst$ is
\begin{eqnarray}
\label{eq:OBSCENTROID}
\hat Z^1_1(\Iobs,\bde_W)&=&\hat Z^1_1(\Iobj,\bde_W)+\hat Z^1_1(\IRCN,\bde_W)=0,
\end{eqnarray}
and by expanding with $\Delta\bth$ and neglecting higher order, we obtain
\begin{eqnarray}
\label{eq:hatZ11}
\hat Z^1_1(\Iobs,\bde_W)&=&\int d^2\theta \lr{\theta-\CSE}^1_1 \Iobjt W(\bth-\bCSE,\bde_W)
+\hat Z^1_1(\IRCN,\bde_W)
\nonumber\\&&\hspace{-100pt}\approx
W(\Delta\bth,\bde_W)\lrs{-\CSE^1_1\lr{Z^0_0-\frac{Z^2_0-\bde_W^*Z^2_2}{\sigma_W^2}}-\CSE^{1*}_1\lr{-\frac{Z^2_2-\bde_WZ^2_0}{\sigma_W^2}}}(\Iobj,\bde_W)
+\hat Z^1_1(\IRCN,\bde_W)
\nonumber\\&&\hspace{-50pt}
\equiv -\frac{W(\Delta\bth,\bde_W)Z^0_0(I^{obj},\bde_W)}{2}\lr{\cC2^0_{0+}\Delta\theta+\cC0^0_{2-}\Delta\theta^*}+\hat Z^1_1(I^{RCN},\bde_W)=0,
\end{eqnarray}
where 
we use the elliptical Gaussian form for the weight function to write the
weight function in the following form
\begin{eqnarray}
W(\bth-\bCSE,\bde_W)&\approx&W(\Delta\bth,\bde_W)\lr{1+\frac{\lr{\CSE^{1*}_1-\bde_W^*\CSE^1_1}\theta^1_1+\lr{\CSE^1_1-\bde_W\CSE^{1*}_1}\theta^{1*}_1}{\sigma_W^2}}W(\bth,\bde_W)\\
W(\Delta\bth,\bde_W)&=& e^{-\frac{\CSE^2_0-\Real{\bde_W^*\CSE^2_2}}{\sigma_W^2}},
\end{eqnarray}
C coefficients are defined as
\begin{eqnarray}
\label{eq:CNM}
\cC^{N }_{M-}&\equiv&\sigma_W^2\lrs{\frac12\lr{\lr{N-M}\cH^{N-2}_M+\lr{N+M-2}\bde_W  \cH^{N-2}_{M-2}}-2\frac{1-\delta^2_W}{\sigma_W^2}\cH^N_M}(I^{obj},Z^N_0,\bde_W)
\\
\cC^{N }_{M+}&\equiv&\sigma_W^2\lrs{\frac12\lr{\lr{N+M}\cH^{N-2}_M+\lr{N-M-2}\bde_W^*\cH^{N-2}_{M+2}}-2\frac{1-\delta^2_W}{\sigma_W^2}\cH^N_M}(I^{obj},Z^N_0,\bde_W)
\\
\label{eq:tCNM}
\cCX^N_{ M-}&\equiv&\lrs{X\cH^N_M-\frac{2}{\sigma_W^2}\lr{\cH^{N+2}_M-\bde  _W\cH^{N+2}_{M-2}}}(I^{obj},Z^N_0,\bde_W)
\\
\cCX^N_{ M+}&\equiv&\lrs{X\cH^N_M-\frac{2}{\sigma_W^2}\lr{\cH^{N+2}_M-\bde^*_W\cH^{N+2}_{M+2}}}(I^{obj},Z^N_0,\bde_W)
\end{eqnarray}
These have non dimension and have spin-M, $X$ is an integer. Here we  
neglect odd order of the complex moments of $\Iobjt$ (i.e. $Z^1_1(\Iobj,\bde_W)=Z^3_M(\Iobj,\bde_W)=0$).
Finally, we obtain $\Delta\bth$ as
\begin{eqnarray}
\label{eq:CSE}
\Delta\bth\approx W(\Delta\bth,\bde_W)\Delta\bth&=&\frac{2}{Z^0_0(I^{obj},\bde_W)}\frac{\lrs{\cC2^0_{0-}\hat Z^1_1-\cC0^0_{2-}\hat Z^{1*}_1}(I^{RCN},\bde_W)}{|\cC2^0_{0+}|^2-|\cC0^0_{2-}|^2}
\nonumber\\
&\equiv&\frac{2}{Z^0_0(I^{obj},\bde_W)}\lrs{\tcC2^0_{0-}\hat Z^1_1-\tcC0^0_{2-}\hat Z^{1*}_1}(I^{RCN},\bde_W),
\end{eqnarray}
where
\begin{eqnarray}
\tcC2^0_{0\pm} &\equiv& \frac{\cC2^0_{0\pm}}{|\cC2^0_{0+}|^2-|\cC0^0_{2-}|^2}\\
\tcC0^0_{2\pm} &\equiv& \frac{\cC0^0_{2\pm}}{|\cC2^0_{0+}|^2-|\cC0^0_{2-}|^2}
\end{eqnarray}
Therefore, $\Delta\bth$ comes from the dipole moments of RCN.

The averaged value of $\Delta\bth$ is 0, because averaged value of $Z^1_1(I^{RCN},\bde_W)$ is also 0,
therefore there is no SCE in averaged sense.
But the averaged value of $(\Delta\bth)^2$ is not 0 and obtained as follows.
\begin{eqnarray}
\overline{ W(\Delta\bth,\bde_W) \Delta\bth}&=&0\\
\label{eq:DG2DT20}
\overline{ W^2(\Delta\bth,\bde_W) \CSE^2_0}&=&\frac{1}{WSN^2}\frac{\sigma_W^2}{1-\delta_W^2}\lr{|\tcC2^0_{0-}|^2+|\tcC0^0_{2-}|^2-2\Real{\tcC2^0_{0-}\tcC0^0_{2-}\bde_W^*}}\\
\label{eq:DG2DT22}
\overline{ W^2(\Delta\bth,\bde_W) \CSE^2_2}&=&\frac{1}{WSN^2}\frac{\sigma_W^2}{1-\delta_W^2}\lr{\lr{\tcC2^0_{0-}}^2\bde_W-2\tcC2^0_{0-}\tcC0^0_{2-}+\lr{\tcC0^0_{2-}}^2\bde_W^*}.
\end{eqnarray}
We define ellipticity of distribution of  CSE as
\begin{eqnarray}
\bde_C &\equiv& \frac{\overline{\CSE^2_2}}{\overline{\CSE^2_0}} \approx\frac{\overline{ W(\Delta\bth,\bde_W)\CSE^2_2}}{\overline{ W(\Delta\bth,\bde_W)\CSE^2_0}}\\
\delta_C &\equiv&\lrabs{\bde_C}.
\end{eqnarray}
It the ellipticity due to CSE distribution.

If the object has an elliptical Gaussian image $I^{EG}(\bth,\bde_I)$
\begin{eqnarray}
I^{EG}(\bth,\bde_I) = Ae^{-\frac{\theta^2_0-\Real{\bde_I^*\theta^2_2}}{\sigma_W^2}}
\end{eqnarray}
where A is an arbitrary amplitude (hereafter EGI) and we use KSB method with Gaussian weight(hereafter KSBGW).
We can obtain eq.(\ref{eq:DG2DT20}) and  eq.(\ref{eq:DG2DT22}) analytically as
\begin{eqnarray}
\label{eq:DG2DT20KSBG}
\overline{\Delta\theta^2_0}\approx\overline{ W^2(\Delta\bth,\bde_W) \Delta\theta^2_0}&=&\frac{\sigma_W^2}{\WSN^2}\lr{1+\frac54\delta^2_I}\\
\label{eq:DG2DT22KSBG}
\overline{\Delta\theta^2_2}\approx\overline{ W^2(\Delta\bth,\bde_W) \Delta\theta^2_2}&=&\frac{\sigma_W^2\bde_I}{\WSN^2}\lr{1+\frac32\delta^2_I}\\
\bde_C&=&\bde_I\frac{1+\frac32\delta_I^2}{1+\frac54\delta_I^2}\approx\lr{1+\frac14\delta_I^2}\bde_I,
\end{eqnarray}
where the detail value of C coefficients in this situation are shown in Appendix \ref{AP:CNM_KSB}.

Fig.\ref{fig:E-HOLICs_3_KSB_C} shows the comparison between the result of simulation and the predicted result using eq.(\ref{eq:DG2DT20KSBG}) and eq.(\ref{eq:DG2DT22KSBG}) with $\bde_I=(0.5, 0)$ and $\sigma_W^2=200$.
We can see these equations are very good approximation except in low WSN.
\begin{figure*}[htbp]	
\epsscale{1.}
\plotone{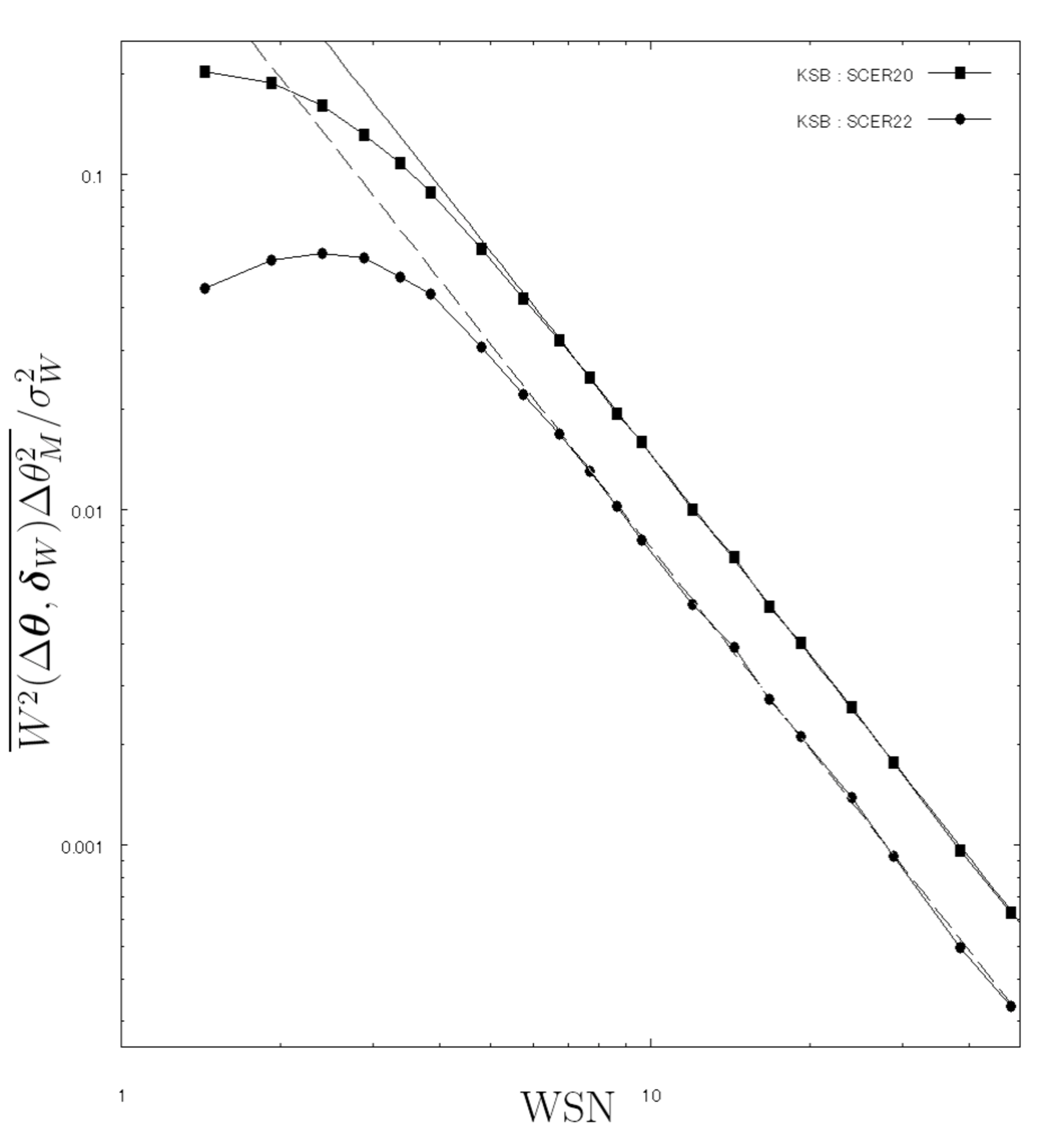}
\caption{
\label{fig:E-HOLICs_3_KSB_C}
Plots are $\overline{ W^2(\Delta\bth,\bde_W) \Delta\theta^2_M}/\sigma_W^2$ in the situation of EGI and KSBGW.
Horizontal axis means WSN and vertical axis means $\overline{ W^2(\Delta\bth,\bde_W) \Delta\theta^2_M}/\sigma_W^2$.
Square(circle) means $\overline{ W^2(\Delta\bth,\bde_W) \Delta\theta^2_0}/\sigma_W^2(\overline{ W^2(\Delta\bth,\bde_W) \Delta\theta^2_2}/\sigma_W^2)$ measured from simulation data and solid(dashed) line means analytical prediction i.e. eq.(\ref{eq:DG2DT20})(eq.(\ref{eq:DG2DT22})). 
} 
\end{figure*}  

\subsection{Complex Moments with Random Count Noise}
In this section, we consider the systematic error in measuring the complex moments due to RCN.

By expanding eq.(\ref{eq:DeltaCM}),
$\hat Z^N_M$ can be expressed as a function of $Z^N_M$, so
\begin{eqnarray}
\label{eq:hatZNM}
\hat Z^N_M(\Iobs,\bde_W)&=&\int d^2\theta \lr{\theta-\Delta\theta}^N_M\Iobst W\lr{\bth-\Delta\bth,\bde_W}
\nonumber\\&=&
\int d^2\theta \lr{\theta-\Delta\theta}^N_M\Iobst W\lr{\bth,\bde_W}W\lr{\Delta\bth,\bde_W}e^{\frac{\lr{\Delta\bth^*-\bde_W^*\Delta\bth}\bth+\lr{\Delta\bth-\bde_W\Delta\bth^*}\bth^*}{\sigma_W^2}}
\nonumber\\&=&
\int d^2\theta \lr{\theta-\Delta\theta}^N_M\lr{\Iobjt+\IRCNt} W\lr{\bth,\bde_W}\lr{1-\frac{\CSE^2_0-\Real{\bde_W^*\CSE^2_2}}{\sigma_W^2}}
\nonumber\\&&\times
\Biggl(1+\frac{\lr{\Delta\bth^*-\bde_W^*\Delta\bth}\bth+\lr{\Delta\bth-\bde_W\Delta\bth^*}\bth^*}{\sigma_W^2}+\frac1{2\sigma_W^4}\lr{\Delta\bth^*-\bde_W^*\Delta\bth}^2\theta^2_2
\nonumber\\&&
+\frac1{2\sigma_W^4}\lr{\Delta\bth-\bde_W\Delta\bth^*}^2\theta^{*2}_2
+\frac1{\sigma_W^4}\lr{\Delta\bth-\bde_W\Delta\bth^*}^2_0\theta^2_0+...\Biggr).
\end{eqnarray}
Here, we define the effects of RCN in measuring the complex moments as
\begin{eqnarray}
\hat Z^N_M(I^{obs},\bde_W)\equiv Z^N_0(I^{obj},\bde_W)\lrs{\cH^N_M(I^{obj},Z^N_0,\bde_W) + \Delta \cH^N_{M(1)}(\bde_W) + \Delta \cH^N_{M(2)}(\bde_W)},
\end{eqnarray}
where $X$  of $\Delta\cH^N_{M(X)}$ means from $X$th order effect of $\IRCNt$.
Because effect from 1st order of $\IRCNt$ is 0, so
\begin{eqnarray}
\overline{\Delta \cH^N_{M(1)}(\bde_W)}&=&0\\
\overline{\Delta \cH^N_{M(2)}(\bde_W)}&\neq&0\\
\overline{\Delta \cH^N_{M(1)}(\bde_W)\Delta Z^O_{P(1)}(\bde_W)}&\neq&0.
\end{eqnarray}
1st and 2nd order of effects for complex moments from $\IRCNt$ can be written as
\begin{eqnarray}
\Delta\cH^N_{M(1)}(\bde_W)=\frac{Z^N_M(I^{RCN},\bde_W)}{Z^N_0(I^{obj},\bde_W)},
\end{eqnarray}
\begin{eqnarray}
\label{eq:CMOM}
\Delta\cH^N_{M(2)}(\bde_W)&\approx& 
\frac{1}{4\sigma_W^2}\Bigglr{\DDT^{2 }_0\lr{\frac{N+M}{\lr{1-\delta^2_W}}\cC^{N}_{M-}-\cC2^N_{M+}}
+\DDT^{2*}_0\lr{\frac{N-M}{\lr{1-\delta^2_W}}\cC^{N}_{M+}-\cC2^N_{M-}}
\nonumber\\&&\hspace{-0pt}
+\DDT^{2 }_2\lr{\frac{N+M}{\lr{1-\delta^2_W}}\cC^{N}_{M-2+}-\cC0^N_{M-2+}}
+\DDT^{2*}_2\lr{\frac{N-M}{\lr{1-\delta^2_W}}\cC^{N}_{M+2-}-\cC0^N_{ M+2-}}}
\nonumber\\&&\hspace{-50pt}
+\frac{2\cH^0_0(I^{obj},Z^N_0,\bde_W)}{WSN^2S_W}
\Bigglr{-\frac12\Biglr{(N+M)\lr{\tcC2^0_{0-}G^N_M-\tcC0^0_{2-}G^N_{M-2}}
\nonumber\\&&\hspace{50pt}
+(N-M)\lr{\tcC2^0_{0+}G^N_M-\tcC0^{0*}_{2-}G^N_{M+2}}}
\nonumber\\&&\hspace{-0pt}
+\frac{1}{\sigma_W^2}\lr{\tcC2^0_{0-}G^{N+2}_M-\tcC0^0_{2-}G^{N+2}_{M-2}-\bde_W  \tcC2^0_{0+}G^{M+2}_{N-2}+\bde_W  \tcC0^{0*}_{2-}G^{N+2}_M}
\nonumber\\&&\hspace{-0pt}
+\frac{1}{\sigma_W^2}\lr{\tcC2^0_{0+}G^{N+2}_M-\tcC0^{0*}_{2-}G^{N+2}_{M+2}-\bde^*_W\tcC2^0_{0-}G^{M+2}_{N+2}+\bde^*_W\tcC0^0_{2-}G^{N+2}_M}},
\end{eqnarray}
where
\begin{eqnarray}
\DDT^2_0&\equiv&\Delta\theta^2_0-\bde_W^*\Delta\theta^2_2\\
\DDT^2_2&\equiv&\Delta\theta^2_2-\bde_W  \Delta\theta^2_0.
\end{eqnarray}
So, we can obtain systematic error ratio "$\overline{\Delta\cH^N_{M(2)}(\bde_W)}$" by C coefficients (i.e. combinations of complex moments).

Here, we define systematic error ratio(hereafter SER) of complex moments as
\begin{eqnarray}
{\rm SER}\equiv\frac{\overline{Z^N_M(I^{obs},\bde_W)}-Z^N_M(I^{obj},\bde_W)}{Z^N_M(I^{obj},\bde_W)}
\end{eqnarray}
and 2nd order of systematic error ratio(hereafter 2ndSER) of complex moments as 
\begin{eqnarray}
{\rm 2ndSER}\equiv\frac{\overline{\Delta\cH^N_{M(2)}(\bde_W)}}{\cH^N_M(I^{obj},Z^N_0,\bde_W)}
\approx\frac{\overline{\Delta\cH^N_{M(2)}(\bde_W)}}{\cH^N_M(I^{obs},Z^N_0,\bde_W)}
\end{eqnarray}
If 2nd order effect of RCN is dominant, these SERs are almost same(SER$\approx$2ndSER).
So, a correction formula for HOLICs which corrects systematic error is defined as
\begin{eqnarray}
\label{eq:cformH}
\cH^{N(corrected)}_M(I^{obj},Z^N_0,\bde_W) &\equiv& \cH^N_M(I^{obj},Z^N_0,\bde_W)-\overline{\Delta\cH^N_{M(2)}(\bde_W)}
\nonumber\\&\approx&
\cH^N_M(I^{obs},Z^N_0,\bde_W)-\overline{\Delta\cH^N_{M(2)}(\bde_W)}.
\end{eqnarray}

\begin{figure*}[htbp]	
\epsscale{1.0}
\plotone{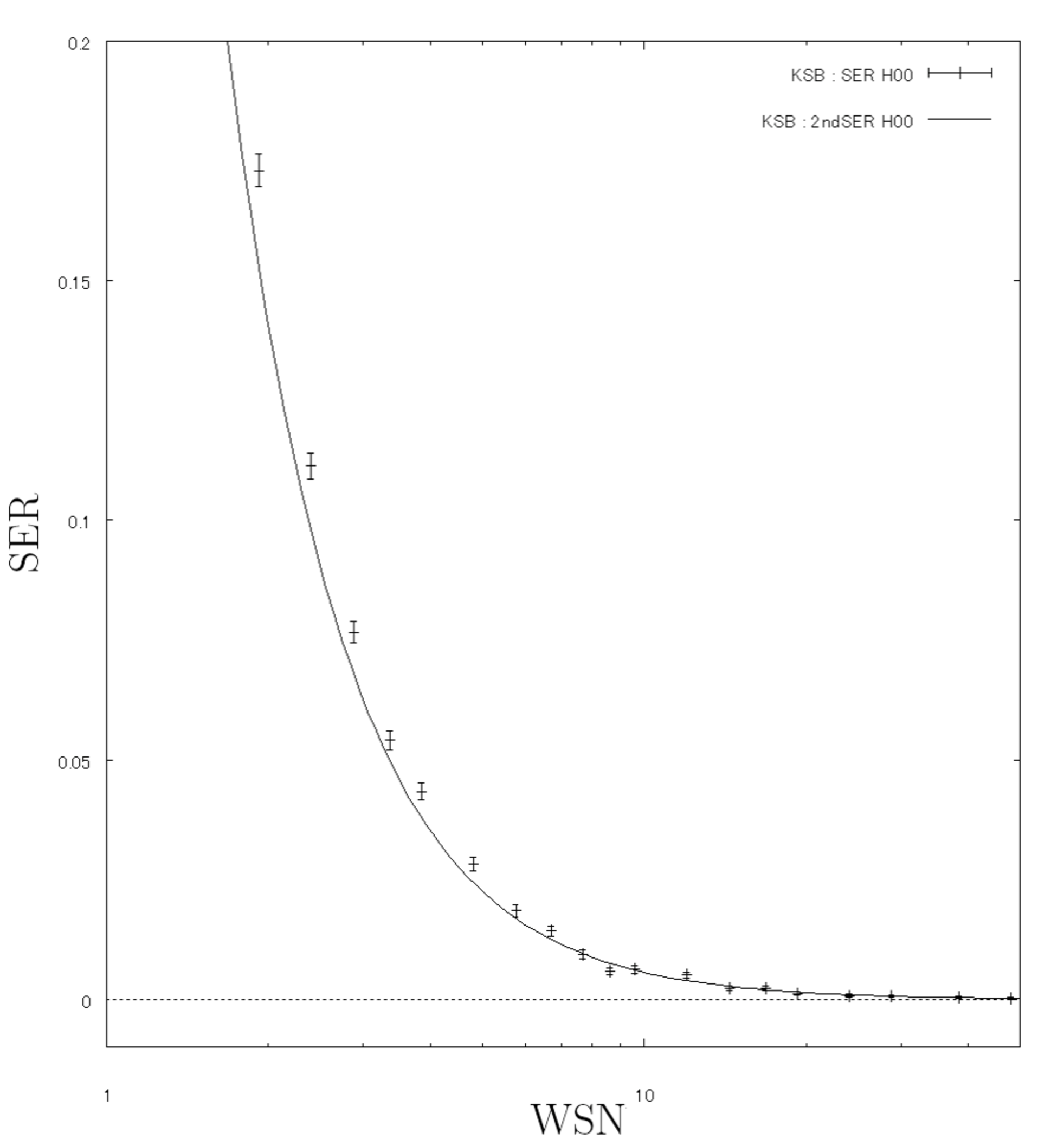}
\caption{
\label{fig:E-HOLICs_3_KSB_H0}
SER due to RCN for monopole moments measured in the situation of EGI and KSBGW.
Horizontal axis means WSN and vertical axis means SER.
Plots are simulated results of SER of $\Delta\cH^0_0$ and line is 2ndSER calculated from eq.(\ref{eq:hatH00GKSB}).
} 
\end{figure*}

\begin{figure*}[htbp]	
\epsscale{1.0}
\plotone{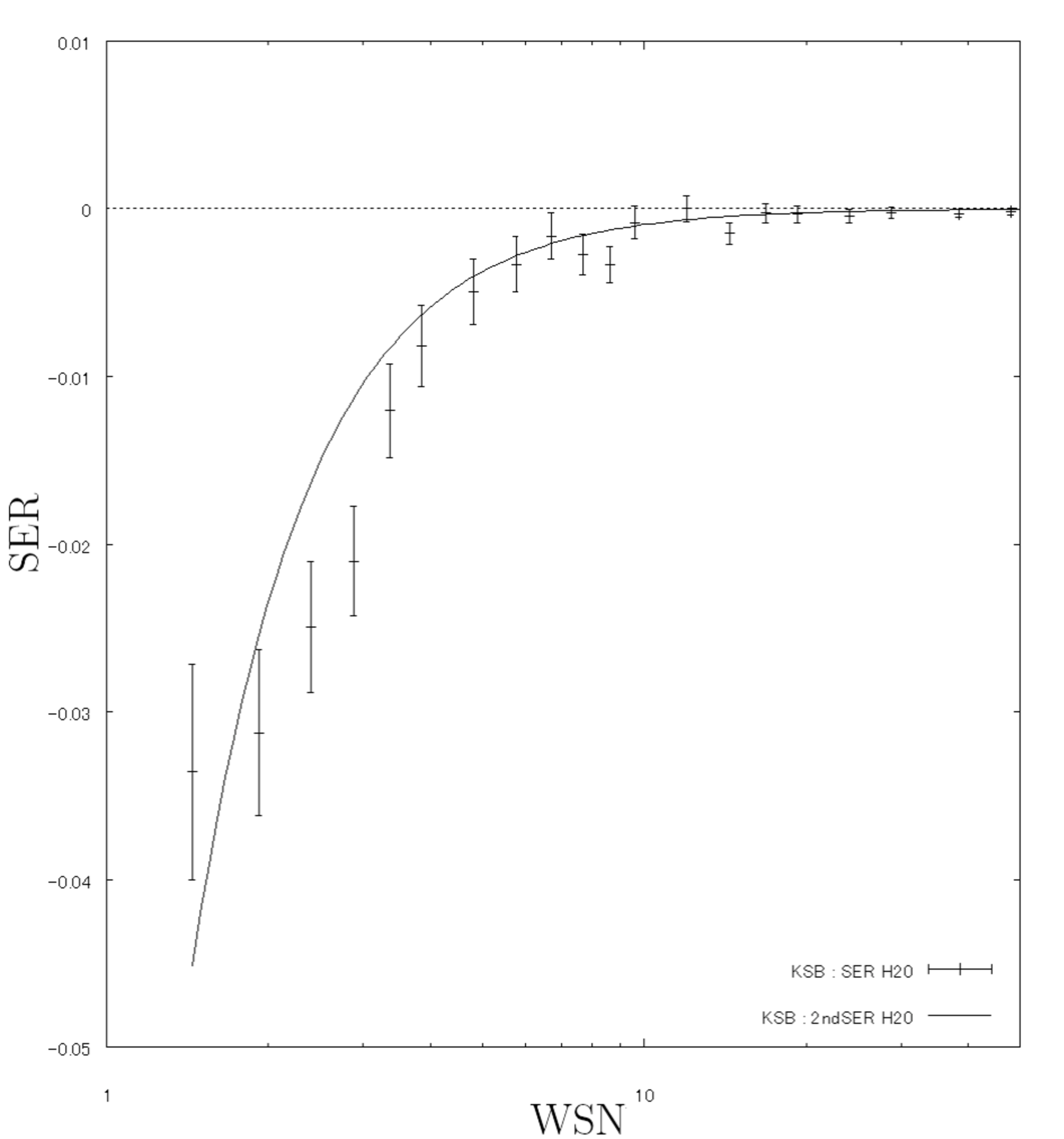}
\caption{
\label{fig:E-HOLICs_3_KSB_H20}
SER due to RCN for $\cH^2_0$ in the situation of EGI and KSBGW.
Horizontal axis means WSN and vertical axis means SER.
Plots are simulated results of SER of $\Delta\cH^2_0$ and line is 2ndSER calculated from eq.(\ref{eq:hatH20GKSB}).
} 
\end{figure*}

\begin{figure*}[htbp]	
\epsscale{1.0}
\plotone{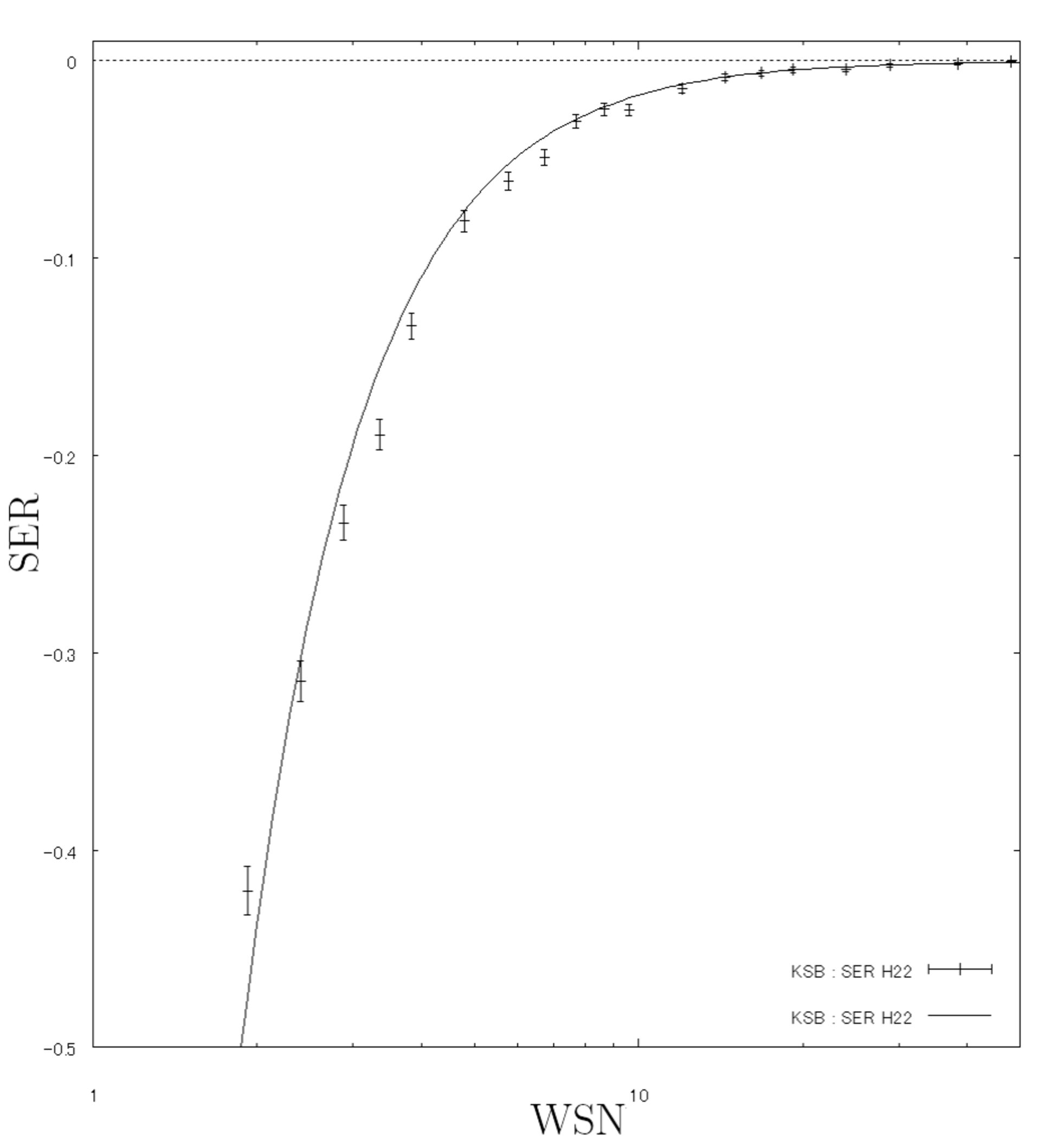}
\caption{
\label{fig:E-HOLICs_3_KSB_H22}
SER due to RCN for $\cH^2_2$ measured in the situation of EGI and KSBGW.
Horizontal axis means WSN and vertical axis means SER.
Plots are simulated results of SER of $\Delta\cH^2_2$ and line is 2ndSER calculated from eq.(\ref{eq:hatH22GKSB}).
} 
\end{figure*}

\begin{figure*}[htbp]	
\epsscale{1.0}
\plotone{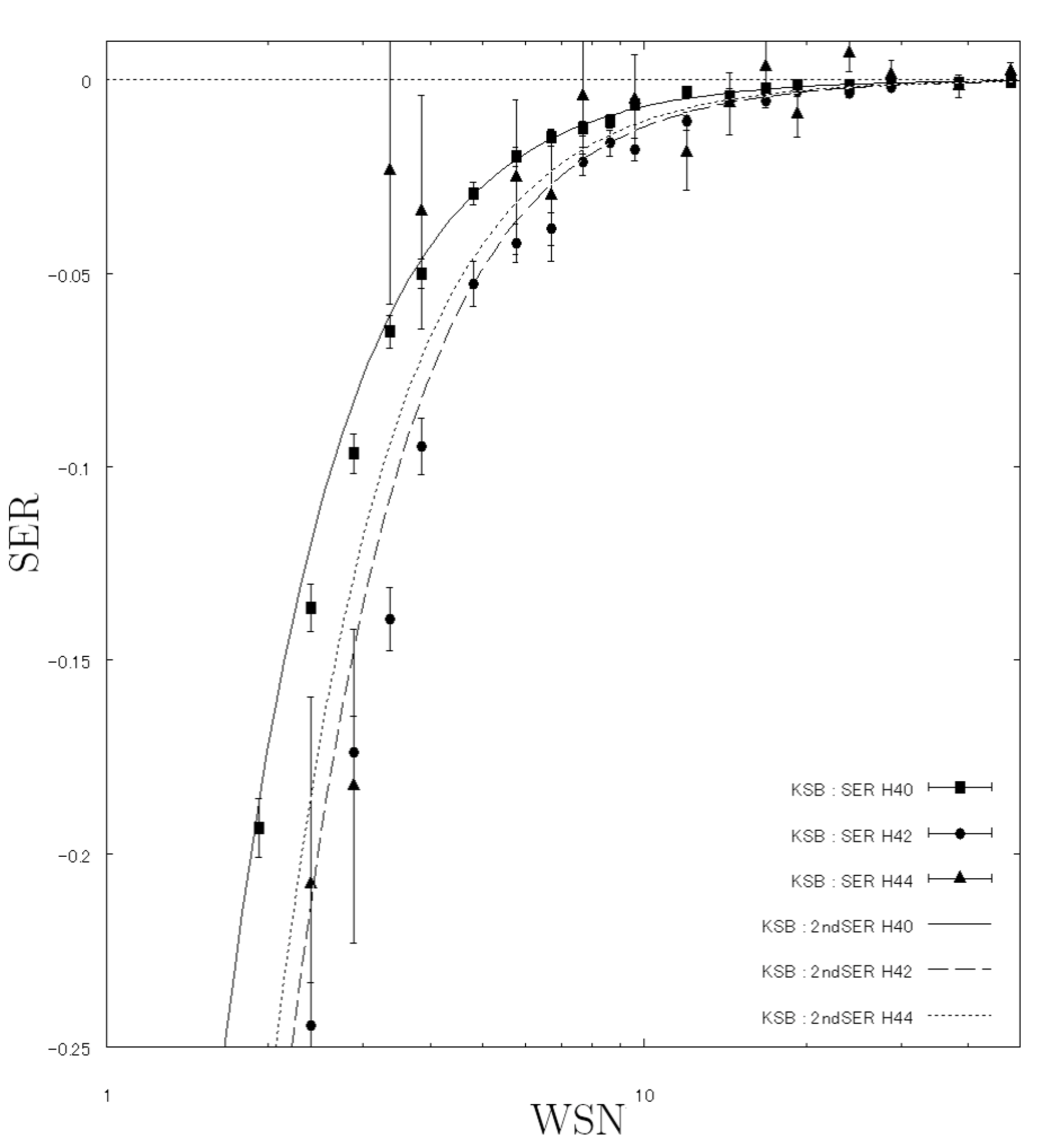}
\caption{
\label{fig:E-HOLICs_3_KSB_H4}
SER due to RCN for $\cH^4_X$ in the situation of EGI and KSBGW.
Horizontal axis means WSN and vertical axis means SER.
Square (circle, triangle) plots are simulated results of $\Delta\cH^4_0$($\Delta\cH^4_2$, $\Delta\cH^4_4$) and solid (long dashed, dashed) line is 2ndSER calculated from eq.(\ref{eq:hatH40GKSB})(eq.(\ref{eq:hatH42GKSB}),eq.(\ref{eq:hatH44GKSB})).
} 
\end{figure*}

In the situation of EGI and KSBGW,
we can obtain the average effects of RCN as
\begin{eqnarray}
\label{eq:hatH00GKSB}
\overline{\Delta\cH^0_{0(2)}(0)}&\approx&
\frac{1}{\WSN^2}\frac{2+\delta^2_I}{4}\cH^0_0(\IEG,Z^0_0,0)\\
\label{eq:hatH20GKSB}
\overline{\Delta\cH^2_{0(2)}(0)}&\approx& 
-\frac{1}{\WSN^2}\frac{3\delta_I^2}{8}\cH^2_0(\IEG,Z^2_0,0)\\
\label{eq:hatH22GKSB}
\overline{\Delta\cH^2_{2(2)}(0)}&\approx& 
-\frac{1}{\WSN^2}\lr{\frac32+\delta_I^2}\cH^2_2(\IEG,Z^2_0,0)
\\
\label{eq:hatH40GKSB}
\overline{\Delta\cH^4_{0(2)}(0)}&\approx& 
-\frac{1}{\WSN^2}\Bigglr{\frac12+\frac34\delta_I^2}\cH^4_0(\IEG,Z^4_0,0)
\\
\label{eq:hatH42GKSB}
\overline{\Delta\cH^4_{2(2)}(0)}&\approx& 
-\frac{1}{\WSN^2}\lr{1+\frac78\delta_I^2}\cH^4_2(\IEG,Z^4_0,0)
\\
\label{eq:hatH44GKSB}
\overline{\Delta\cH^4_{4(2)}(0)}&\approx& 
-\frac{1}{\WSN^2}\Bigglr{\frac12+\frac94\delta_I^2}\cH^4_4(\IEG,Z^4_0,0).
\end{eqnarray}
Fig.\ref{fig:E-HOLICs_3_KSB_H0} to Fig.\ref{fig:E-HOLICs_3_KSB_H4} show simulated results of SER and 2ndSER calculated from these equations.
Fig.\ref{fig:E-HOLICs_3_KSB_H0} shows about monopole moments and eq.(\ref{eq:hatH00GKSB}),
Fig.\ref{fig:E-HOLICs_3_KSB_H20} shows about spin-0 quadrupole moments and eq.(\ref{eq:hatH20GKSB}),
Fig.\ref{fig:E-HOLICs_3_KSB_H22} shows about spin-2 quadrupole moments and eq.(\ref{eq:hatH22GKSB}),
Fig.\ref{fig:E-HOLICs_3_KSB_H4} shows about 16pole moments and eq.(\ref{eq:hatH40GKSB}).
\subsection{Ellipticity with Random Count Noise}
Now we show the systematic error in measuring of ellipticity due to RCN which is 
used in the weak lensing analysis. 
The observed ellipticity and object ellipticity are defined as 
\begin{eqnarray}
\bde_{obs}&=&\hat \cH^2_2(\Iobs,Z^2_0,\bde_W) = \frac{\hat Z^2_2(\Iobs,\bde_W)}{\hat Z^2_0(\Iobs,\bde_W)}\\
\bde_{obj}&=&\cH^2_2(\Iobj,Z^2_0,\bde_W) = \frac{Z^2_2(\Iobj,\bde_W)}{Z^2_0(\Iobj,\bde_W)},
\end{eqnarray}
and 
we consider effect from RCN until 2nd order as
\begin{eqnarray}
\bde_{obs}\approx\bde_{obj} + \Delta\bde_{(1)} + \Delta\bde_{(2)}
\end{eqnarray}
where number of sub scripts mean order of RCN effect.
Here, we define systematic error ratio(hereafter SER) of elliticity as
\begin{eqnarray}
{\rm SER}\equiv\frac{\overline{\bde_{obs}}-\bde_{obj}}{\bde_{obj}}
\end{eqnarray}
and 2nd order of systematic error ratio(hereafter 2ndSER) of ellipticity as 
\begin{eqnarray}
{\rm 2ndSER}\equiv\frac{\overline{\Delta\bde_{(2)}}}{\bde_{obj}}.
\end{eqnarray}
If 2nd order effect of RCN is dominant, these SERs are almost same(SER$\approx$2ndSER)
So, a correction formula for ellipticity is defined as
\begin{eqnarray}
\label{eq:cformD}
\bde^{obj(corrected)}\equiv\bde^{obs}-\overline{\Delta\bde_{(2)}}
\end{eqnarray}

For such combinations of complex moments, we cannot neglect the 1st order effects
because there are combinations of the 1st order effects.
For example, RCN effects for the observed ellipticity can be calculated as 
\begin{eqnarray}
\label{eq:RealE}
\bde_{obs}&=&\frac{\hat Z^2_2(I^{obs},\bde_W)}{\hat Z^2_0(I^{obs},\bde_W)}\approx
\frac{\cH^2_2(I^{obj},\bde_W)+\Delta\cH^2_{2(1)}(\bde_W)+\Delta\cH^2_{2(2)}(\bde_W)}{1+\Delta\cH^2_{0(1)}(\bde_W)+\Delta\cH^2_{0(2)}(\bde_W)}
\nonumber\\&\approx&
\bde_{obj}
+\Delta\cH^2_{2(1)}(\bde_W)+\Delta\cH^2_{2(2)}(\bde_W)
-\bde_{obj}\Delta\cH^2_{0(1)}(\bde_W)
-\bde_{obj}\Delta\cH^2_{0(2)}(\bde_W)
\nonumber\\&&
-\frac{\lr{\cH^0_0(\Iobj,Z^2_0,\bde_W)}^2}{WSN^2}\frac{\lr{G^4_2-\bde_WG^4_0}}{S_W}
\end{eqnarray}
so
\begin{eqnarray}
\Delta\bde_{(1)}&=&
\Delta\cH^2_{2(1)}(\bde_W)
-\bde_{obj}\Delta\cH^2_{0(1)}(\bde_W)\\
\Delta\bde_{(2)}&=&
\Delta\cH^2_{2(2)}(\bde_W)
-\bde_{obj}\Delta\cH^2_{0(2)}(\bde_W)
-\frac{\lr{\cH^0_0(\Iobj,Z^2_0,\bde_W)}^2}{\WSN^2}\frac{G^4_2-\bde_WG^4_0}{S_W}
\end{eqnarray}

\begin{figure*}[htbp]	
\epsscale{1.0}
\plotone{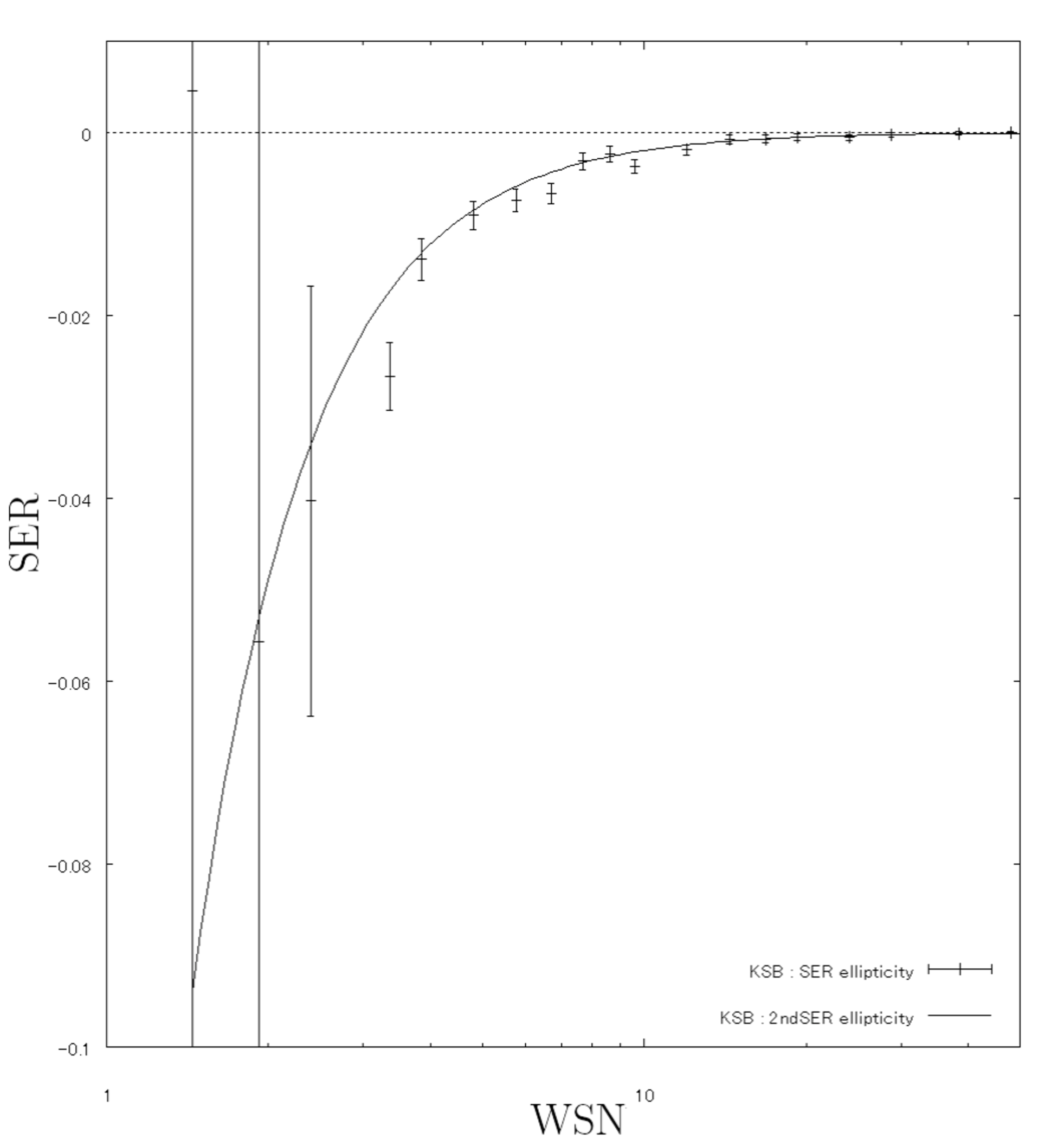}
\caption{
\label{fig:E-HOLICs_3_KSB_EReal}
SER due to RCN for observed ellipticity in the situation of  EGI and KSBGW.
Horizontal axis means WSN and vertical axis means SER $\overline{\Delta \bde_{(2)}}/\bde_{}$.
Plots are simulated results and line is 2ndSER calculated from eq.(\ref{eq:KSB_DD2}).
} 
\end{figure*}

In the situation of EGI and KSBGW, the 
averaged value of eq.(\ref{eq:RealE}) can be calculated analytically as follows.
\begin{eqnarray}
\label{eq:RealEGKSB}
\overline{\bde_{obs}}
&\approx&\bde_{obj}\lr{1-\frac{1}{\WSN^2}\lr{\frac12+\frac{9}{8}\delta_I^2}}\\
\label{eq:KSB_DD2}
\overline{\Delta\bde_{(2)}}&=&-\frac{\bde_{obj}}{\WSN^2}\lr{\frac12+\frac{9}{8}\delta_I^2}.
\end{eqnarray}
Fig.\ref{fig:E-HOLICs_3_KSB_EReal} is the comparison between the result of simulation and prediction 
by eq.(\ref{eq:KSB_DD2}) with $\bde_I=(0.5, 0)$
and in this test we use the same size of image for weight function, 
so $\bde_{obj}=(0.25, 0)$.
We can see eq.(\ref{eq:KSB_DD2}) gives a good agreement except for the sources with low WSN,
and the ellipticity of object having $WSN=5$ is measured with about $4\%$ underestimation.

\section{E-HOLICs method with Random Count Noise}
E-HOLICs method uses the ellipticity of objects for the ellipticity of the weight function. 
However we cannot measure true ellipticity due to random count noise(RCN).
Therefore we must consider RCN effect for the ellipticity of weight function.

\subsection{E-HOLICs method with Random Count Noise and true ellipticity for weight function}
In this section, before treating the realistic situation,
we consider the ideal situation where we use true ellipticity for the weight function. Thus the ellipticity is written as follows: 
\begin{eqnarray}
\bde_{obs}&=&\frac{\hat Z^2_2(\Iobs,\bde_I)}{\hat Z^2_0(\Iobs,\bde_I)}\\
\bde_{obj}&=&\frac{Z^2_2(\Iobj,\bde_W)}{Z^2_0(\Iobj,\bde_W)}=\frac{Z^2_2(\Iobj,\bde_I)}{Z^2_0(\Iobj,\bde_I)}=\bde_W=\bde_I
\end{eqnarray}
Calculations in this section are useful for consideration for the realistic situation which is explained in 
the next section.
C coefficients of ellipticai Gaussian image(EGI) are calculated analytically and shown in Appendix \ref{AP:CNM_TEH}.

Centroid shift error(CSE) for the image with arbitrary distribution is given by eq.(\ref{eq:CSE}), 
and average effects are given by eq.(\ref{eq:DG2DT20}) and eq.(\ref{eq:DG2DT22}).
\begin{eqnarray}
\label{eq:CSEEH}
\Delta\bth\approx W(\Delta\bth,\bde_W)\Delta\bth&=&\frac{2}{Z^0_0(I^{obj},\bde_I)}\lrs{\tcC2^0_{0-}\hat Z^1_1-\tcC0^0_{2-}\hat Z^{1*}_1}(I^{RCN},\bde_I),
\end{eqnarray}
\begin{eqnarray}
\label{eq:DG2DT20RH}
\overline{\CSE^2_0}\approx\overline{ W^2(\Delta\bth,\bde_W) \CSE^2_0}&=&\frac{1}{WSN^2}\frac{\sigma_W^2}{1-\delta_W^2}\lr{|\tcC2^0_{0-}|^2+|\tcC0^0_{2-}|^2-2\Real{\tcC2^0_{0-}\tcC0^0_{2-}\bde_I^*}}\\
\label{eq:DG2DT22EH}
\overline{\CSE^2_2}\approx\overline{ W^2(\Delta\bth,\bde_W) \CSE^2_2}&=&\frac{1}{WSN^2}\frac{\sigma_W^2}{1-\delta_W^2}\lr{\lr{\tcC2^0_{0-}}^2\bde_I-2\tcC2^0_{0-}\tcC0^0_{2-}+\lr{\tcC0^0_{2-}}^2\bde_I^*}.
\end{eqnarray}

\begin{figure*}[htbp]	
\epsscale{1.0}
\plotone{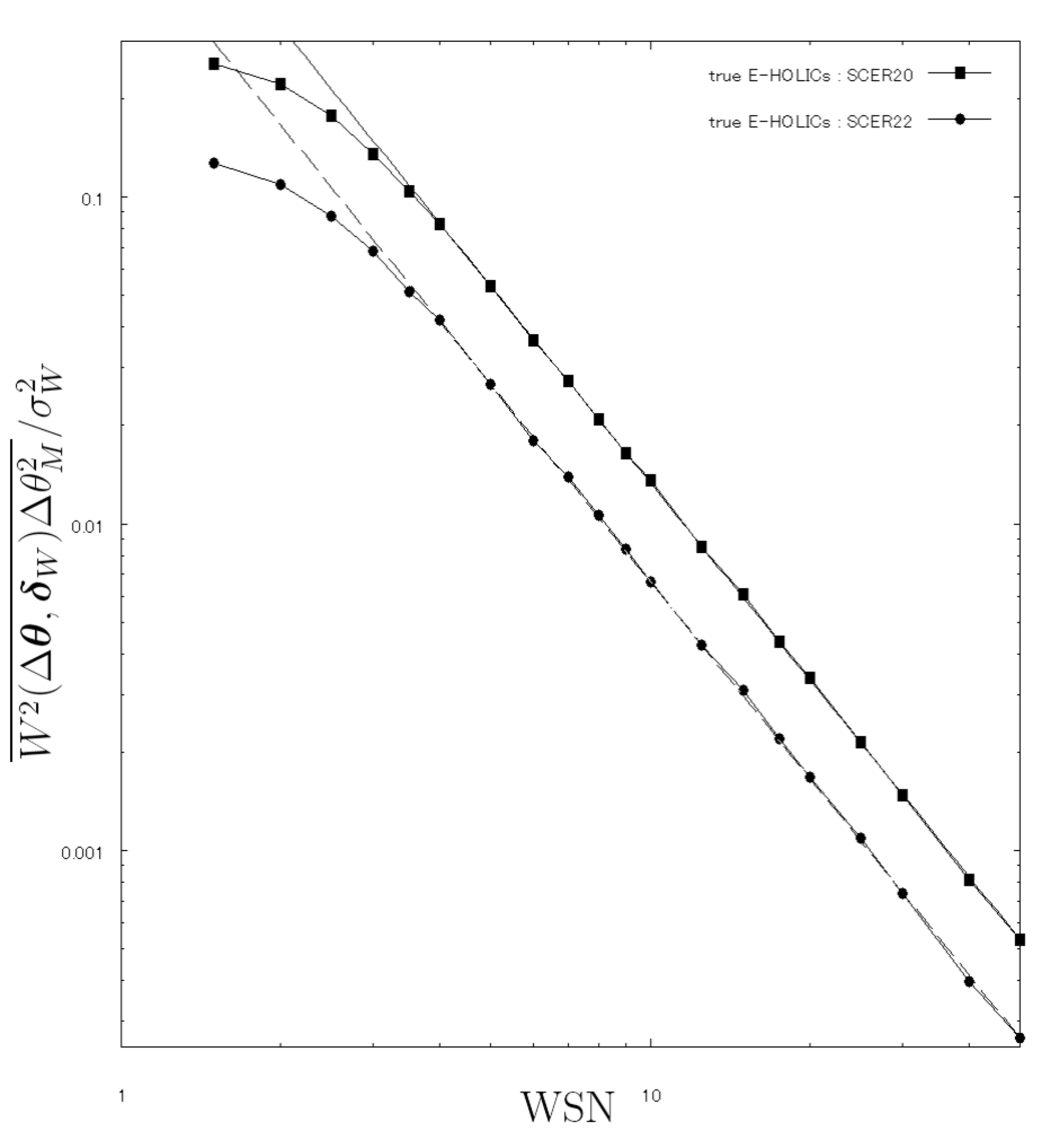}
\caption{
\label{fig:E-HOLICs_3_FIX_C}
Plots are $\overline{\Delta\theta^2_M}/\sigma_W^2$ with elliptical Gaussian image $\bde_I=(0.5, 0)$ and E-HOLICs method $\bde_W=\bde_I$.
Horizontal axis means WSN and vertical axis means $\overline{\Delta\theta^2_M}/\sigma_W^2$.
Square(circle) means $\overline{\Delta\theta^2_0}/\sigma_W^2(\overline{\Delta\theta^2_2}/\sigma_W^2)$ measured from simulation data and solid(dashed) line means analytical prediction i.e. eq.(\ref{eq:DG2DT20Gauss})(eq.(\ref{eq:DG2DT22Gauss})).
} 
\end{figure*}

The average CSE effects of EGI can be obtained as 
\begin{eqnarray}
\label{eq:DG2DT20Gauss}
\overline{\CSE^2_0}\approx\overline{ W^2(\Delta\bth,\bde_W)\CSE^2_0}&=&\frac{1       }{WSN^2}\frac{\sigma^2_W}{1-\delta^2_I}\\
\label{eq:DG2DT22Gauss}
\overline{\CSE^2_2}\approx\overline{ W^2(\Delta\bth,\bde_W)\CSE^2_2}&=&\frac{\bde_I}{WSN^2}\frac{\sigma^2_W}{1-\delta^2_I}\\
\bde_C&=&\bde_I.
\end{eqnarray}
Fig.\ref{fig:E-HOLICs_3_FIX_C} shows the comparison between the results of simulation and the prediction be eq.(\ref{eq:DG2DT20Gauss}) and eq.(\ref{eq:DG2DT22Gauss}) with $\bde_I=(0.5, 0)$.

\begin{figure*}[htbp]	
\epsscale{1.0}
\plotone{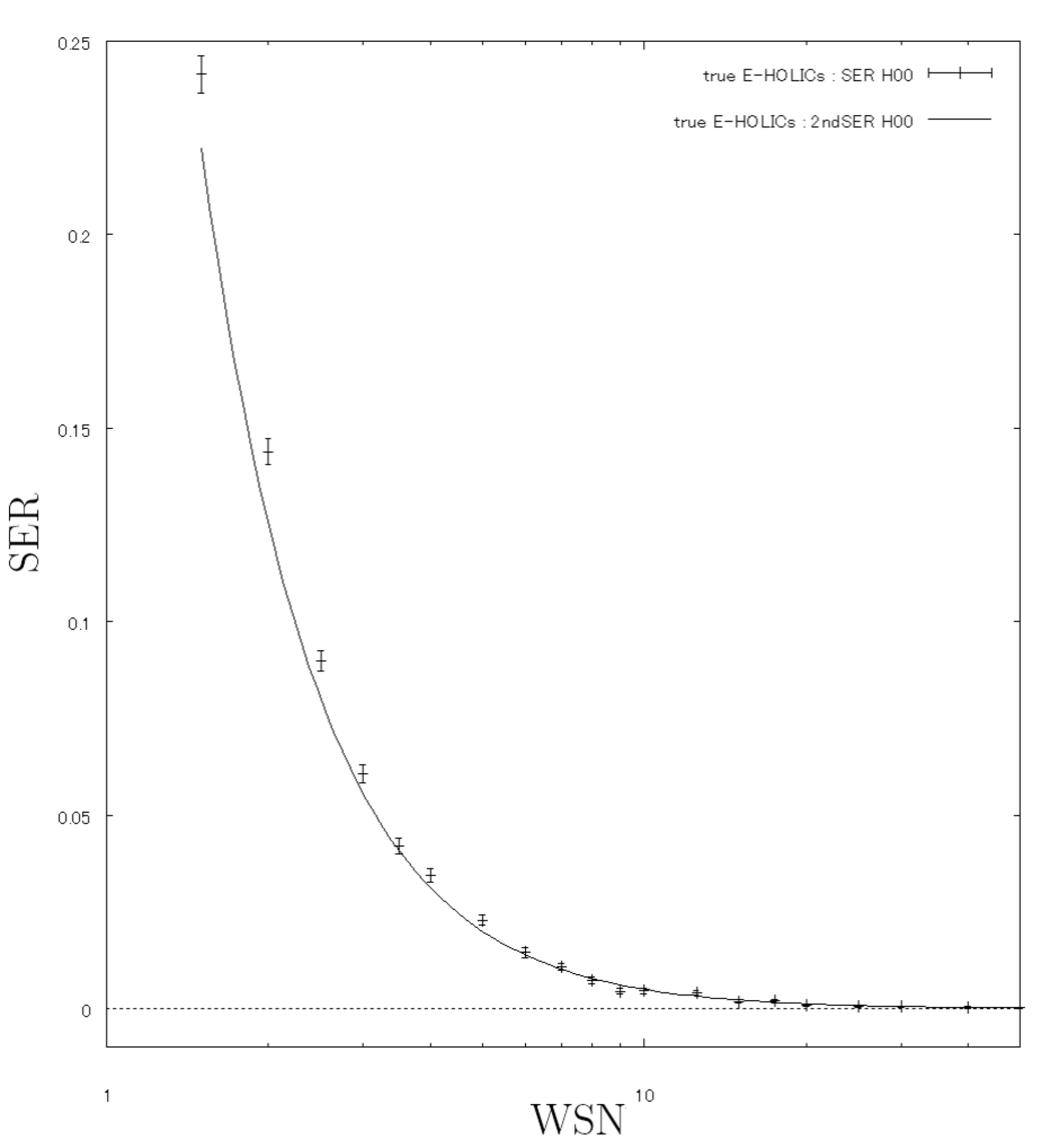}
\caption{
\label{fig:E-HOLICs_3_FIX_H0}
SER due to RCN for monopole moments measured with EGI and E-HOLICs method and true ellipticity for weight function.
Horizontal axis means WSN and vertical axis means systematic error.
Plots are simulated results of SER of $\Delta\cH^0_0$ and line is 2ndSER calculated from eq.(\ref{eq:hatH00EGIEGW}).
} 
\end{figure*}

\begin{figure*}[htbp]	
\epsscale{1.0}
\plotone{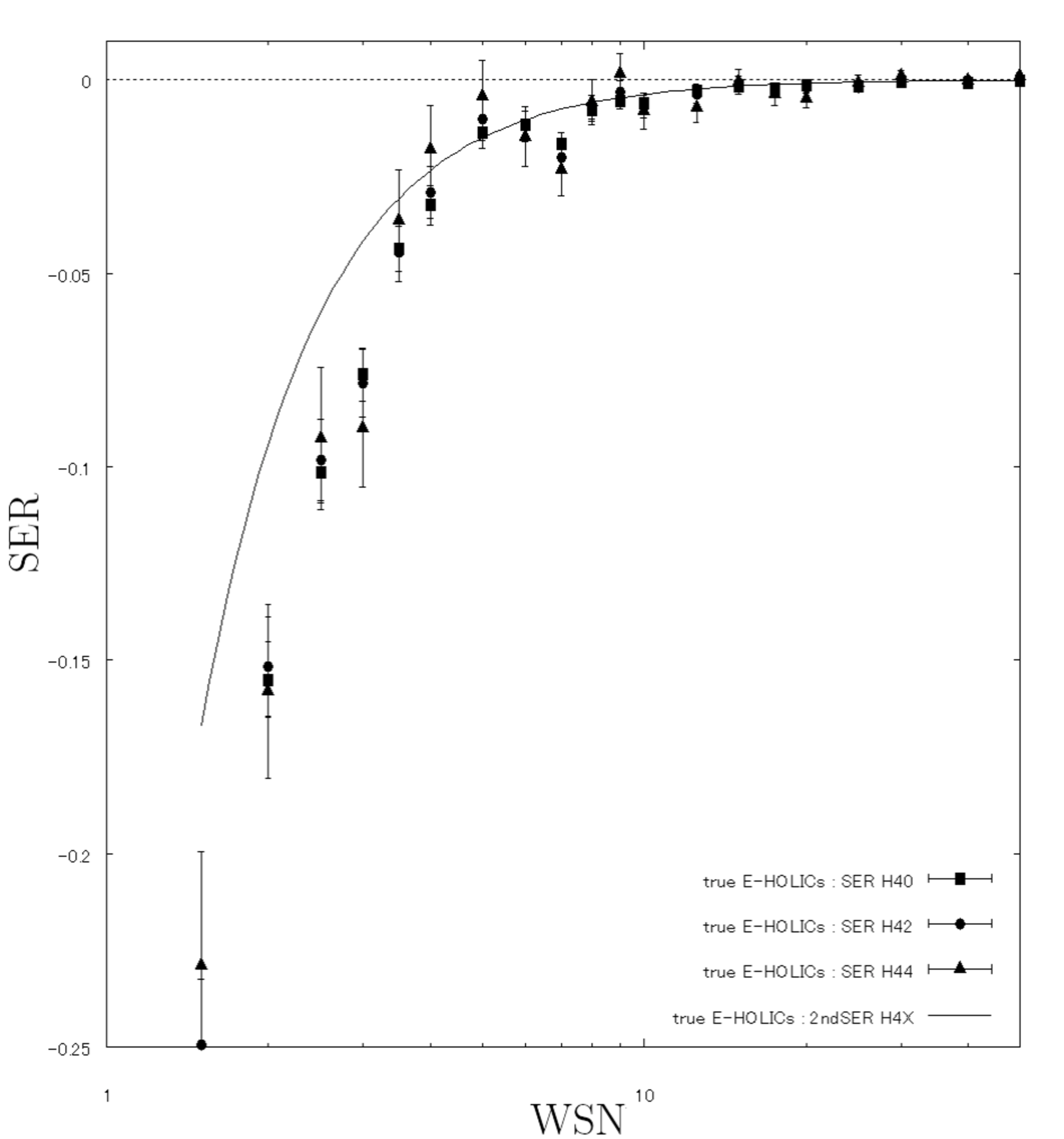}
\caption{
\label{fig:E-HOLICs_3_FIX_H4}
SER due to RCN for quadrupole moments measured with EGI and E-HOLICs method and true ellipticity for weight function.
Horizontal axis means WSN and vertical axis means systematic error.
Square(circle, triangle) plots are simulated results of SER of $\Delta\cH^4_0$($\Delta\cH^4_2$, $\Delta\cH^4_4$) and line is 2ndSER calculated from eq.(\ref{eq:hatH40EGIEGW}), eq.(\ref{eq:hatH42EGIEGW}), eq.(\ref{eq:hatH44EGIEGW}).
} 
\end{figure*}

From eq.(\ref{eq:CMOM}) $\Delta\cH^N_{M(2)}$ are obtained as
\begin{eqnarray}
\label{eq:CMOMEH}
\Delta\cH^N_{M(2)}(\bde_I)&\approx& 
\frac{1}{4\sigma_W^2}\Bigglr{\DDT^{2 }_0\lr{\frac{N+M}{\lr{1-\delta^2_I}}\cC^{N}_{M-}-\cC2^N_{M+}}
+\DDT^{2*}_0\lr{\frac{N-M}{\lr{1-\delta^2_I}}\cC^{N}_{M+}-\cC2^N_{M-}}
\nonumber\\&&\hspace{-0pt}
+\DDT^{2 }_2\lr{\frac{N+M}{\lr{1-\delta^2_I}}\cC^{N}_{M-2+}-\cC0^N_{M-2+}}
+\DDT^{2*}_2\lr{\frac{N-M}{\lr{1-\delta^2_I}}\cC^{N}_{M+2-}-\cC0^N_{ M+2-}}}
\nonumber\\&&\hspace{-10pt}
+\frac{2\cH^0_0(I^{obj},Z^N_0,\bde_I)}{WSN^2S_W}
\Bigglr{-\frac12\Biglr{(N+M)\lr{\tcC2^0_{0-}G^N_M-\tcC0^0_{2-}G^N_{M-2}}
\nonumber\\&&\hspace{50pt}
+(N-M)\lr{\tcC2^0_{0+}G^N_M-\tcC0^{0*}_{2-}G^N_{M+2}}}
\nonumber\\&&\hspace{-0pt}
+\frac{1}{\sigma_W^2}\lr{\tcC2^0_{0-}G^{N+2}_M-\tcC0^0_{2-}G^{N+2}_{M-2}-\bde_I  \tcC2^0_{0+}G^{N+2}_{M-2}+\bde_I  \tcC0^{0*}_{2-}G^{N+2}_M}
\nonumber\\&&\hspace{-0pt}
+\frac{1}{\sigma_W^2}\lr{\tcC2^0_{0+}G^{N+2}_M-\tcC0^{0*}_{2-}G^{N+2}_{M+2}-\bde^*_I\tcC2^0_{0-}G^{N+2}_{M+2}+\bde^*_I\tcC0^0_{2-}G^{N+2}_M}},
\end{eqnarray}
The average effects for the complex moments of EGI can be obtained as 
\begin{eqnarray}
\label{eq:hatH00EGIEGW}
\overline{\Delta\cH^0_{0(2)}(\bde_I)}&\approx&
\frac{1}{2\WSN^2}\cH^0_0(\IEG,Z^0_0,\bde_I)\\
\label{eq:hatH20EGIEGW}
\overline{\Delta\cH^2_{0(2)}(\bde_I)}&\approx& 
0\\
\label{eq:hatH22EGIEGW}
\overline{\Delta\cH^2_{2(2)}(\bde_I)}&\approx& 
0\\
\label{eq:hatH40EGIEGW}
\overline{\Delta\cH^4_{0(2)}(\bde_I)}&\approx& 
-\frac{1}{2\WSN^2}\cH^4_0(\IEG,Z^4_0,\bde_I)\\
\label{eq:hatH42EGIEGW}
\overline{\Delta\cH^4_{2(2)}(\bde_I)}&\approx& 
-\frac{1}{2\WSN^2}\cH^4_2(\IEG,Z^4_0,\bde_I)\\
\label{eq:hatH44EGIEGW}
\overline{\Delta\cH^4_{4(2)}(\bde_I)}&\approx&
-\frac{1}{2\WSN^2}\cH^4_4(\IEG,Z^4_0,\bde_I).
\end{eqnarray}
Fig.\ref{fig:E-HOLICs_3_FIX_H0} to fig.\ref{fig:E-HOLICs_3_FIX_H4} are the comparison between numerical results of SER and theoretical predictions of 2ndSER calculated from eq.(\ref{eq:hatH00EGIEGW}) to eq.(\ref{eq:hatH44EGIEGW}), respectively.  
All cases we assume $\bde_I=(0.5,0)$.
\begin{figure*}[htbp]	
\epsscale{1.0}
\plotone{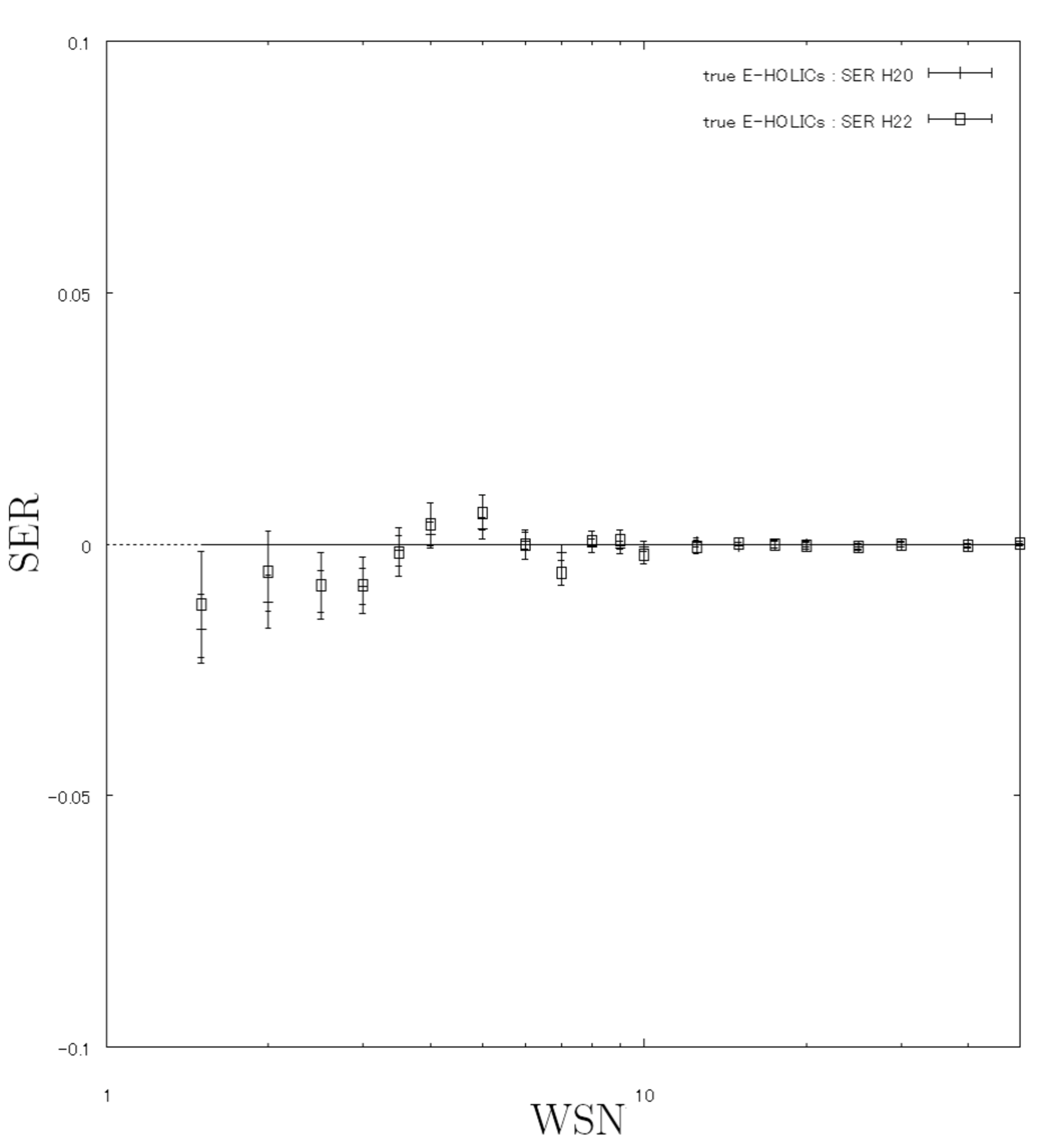}
\caption{
\label{fig:E-HOLICs_3_FIX_H2}
SER due to RCN for quadrupole moments measured with EGI and E-HOLICs method and true ellipticity for weight function.
Horizontal axis means WSN and vertical axis means systematic error.
Cross(square) plots are simulated results of SER of $\Delta\cH^2_0$($\Delta\cH^2_2$).
} 
\end{figure*}  

From eq.(\ref{eq:RealE}) $\overline{\Delta\bde_{(2)}}$ is obtained as
\begin{eqnarray}
\label{eq:RealEtrue}
\overline{\Delta\bde_{(2)}}=
\overline{\Delta\cH^2_{2(2)}(\bde_I)}-\bde_{obj}\overline{\Delta\cH^2_{0(2)}(\bde_I)}
-\frac{\lr{\cH^0_0(\Iobs,Z^2_0,\bde_I)}^2}{WSN^2}\frac{\overline{G^4_2}-\bde_I\overline{G^4_0}}{S_W}
\end{eqnarray}

In the situation of EGI, the systematic error is obtained as
\begin{eqnarray}
\label{eq:RealEGI}
\overline{\bde_{obs}}
&=&\bde_I\lr{1-\frac{1}{WSN^2}\frac{1-\delta_I^2}{2}}\\
\label{eq:FIX_DD2}
\overline{\Delta\bde_{(2)}}&=&-\frac{\bde_I}{WSN^2}\frac{1-\delta_I^2}{2}.
\end{eqnarray}

\begin{figure*}[htbp]	
\epsscale{1.0}
\plotone{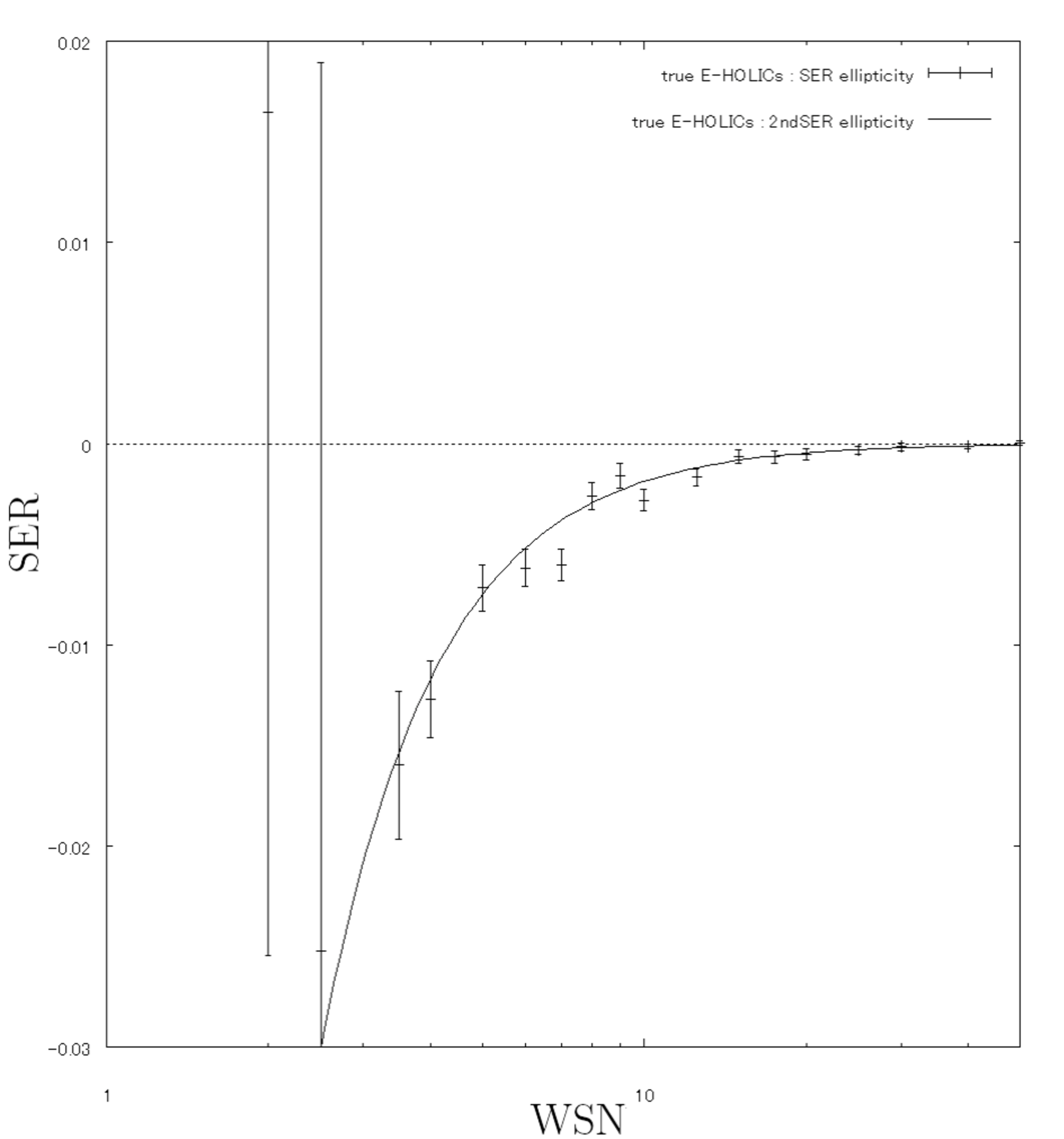}
\caption{
\label{fig:E-HOLICs_3_FIX_EReal}
SER due to RCN for observed ellipticity measured with EGI and E-HOLICs method with true ellipticity for weight function $\bde_W=\bde_I=(0.5, 0)$.
Horizontal axis means WSN and vertical axis means SER $\overline{\Delta\bde_{(2)}}/\bde_I$.
Plots are simulated results and line is 2ndSER calculated from eq.(\ref{eq:FIX_DD2}).
} 
\end{figure*}

Fig.\ref{fig:E-HOLICs_3_FIX_EReal} shows the comparison between numerical results of SER and theoretical predation of 2ndSER given by eq.(\ref{eq:RealEGI}) with $\bde_I=(0.5,0)$.
The ellipticity is defined by quadrupole moments and 
the averaged effect for quadrupole moments is 0 (eq.(\ref{eq:hatH20EGIEGW}) and eq.(\ref{eq:hatH22EGIEGW})),
however the ellipticity has non-zero average. It comes from the combination of 1st order effect in the complex moments due to $\IRCNt$.
In this situation, we observe that the ellipticity is underestimated by $1.5\%$ in average if $WSN=5$.

We can see the these equations give reasonably good fitting formulas except for the images with low WSN.

\subsection{E-HOLICS method with Random Count Noise and without PSF Correction}
In the previous section, we derived correction formulas for the systematic error in E-HOLICs method 
in the case of ideal situation, namely we used true ellipticity. 
In the application of E-HOLICs method for realistic situation, we cannot use ellipticity of object 
because of f RCN effect.
Thus in reality we define the ellipticities as follows.
\begin{eqnarray}
\bde_{obj}&=&\frac{Z^2_2(\Iobj,\bde_I)}{Z^2_0(\Iobj,\bde_I)}=\bde_I\\
\bde_{obs}&=&\frac{\hat Z^2_2(\Iobs,\bde_W)}{\hat Z^2_0(\Iobs,\bde_W)}=\bde_W
\end{eqnarray}
Now we consider RCN effects in the ellipticity for weight function.

\subsubsection{Centroid Shift Error}
CSE $\Delta\bth$ is already 1st order of $\IRCNt$
and we use square of CSE(i.e. $\Delta\theta^2_0$) in calculations of the complex moments.
Therefore we may be able to neglect the differences between $\bde_I$ and $\bde_W$ in the calculation of  CSE 
because we expect these are higher order effects,
Thus  we obtain
\begin{eqnarray}
\Delta\bth\approx W(\Delta\bth,\bde_W)\Delta\bth&=&\frac{2}{Z^0_0(I^{obj},\bde_W)}\lrs{\tcC2^0_{0-}\hat Z^1_1-\tcC0^0_{2-}\hat Z^{1*}_1}(I^{RCN},\bde_W)
\nonumber\\&\approx&
\frac{2}{Z^0_0(I^{obj},\bde_I)}\lrs{\tcC2^0_{0-}\hat Z^1_1-\tcC0^0_{2-}\hat Z^{1*}_1}(I^{RCN},\bde_I)
\end{eqnarray}
and the averaged value as
\begin{eqnarray}
\overline{\Delta\bth}\approx\overline{ W(\Delta\bth,\bde_W) \Delta\bth}&=&0\\
\label{eq:DG2DT20EH}
\overline{\Delta\theta^2_0}\approx\overline{ W^2(\Delta\bth,\bde_W) \Delta\theta^2_0}&\approx&\frac{1}{WSN^2}\frac{\sigma_W^2}{1-\delta_I^2}\lr{|\tcC2^0_{0-}|^2+|\tcC0^0_{2-}|^2-2\Real{\tcC2^0_{0-}\tcC0^0_{2-}\bde_I^*}}\\
\label{eq:DG2DT22EH}
\overline{\Delta\theta^2_2}\approx\overline{ W^2(\Delta\bth,\bde_W) \Delta\theta^2_2}&\approx&\frac{1}{WSN^2}\frac{\sigma_W^2}{1-\delta_I^2}\lr{\lr{\tcC2^0_{0-}}^2\bde_I-2\tcC2^0_{0-}\tcC0^0_{2-}+\lr{\tcC0^0_{2-}}^2\bde_I^*}
\end{eqnarray}

\begin{figure*}[htbp]	
\epsscale{1.0}
\plotone{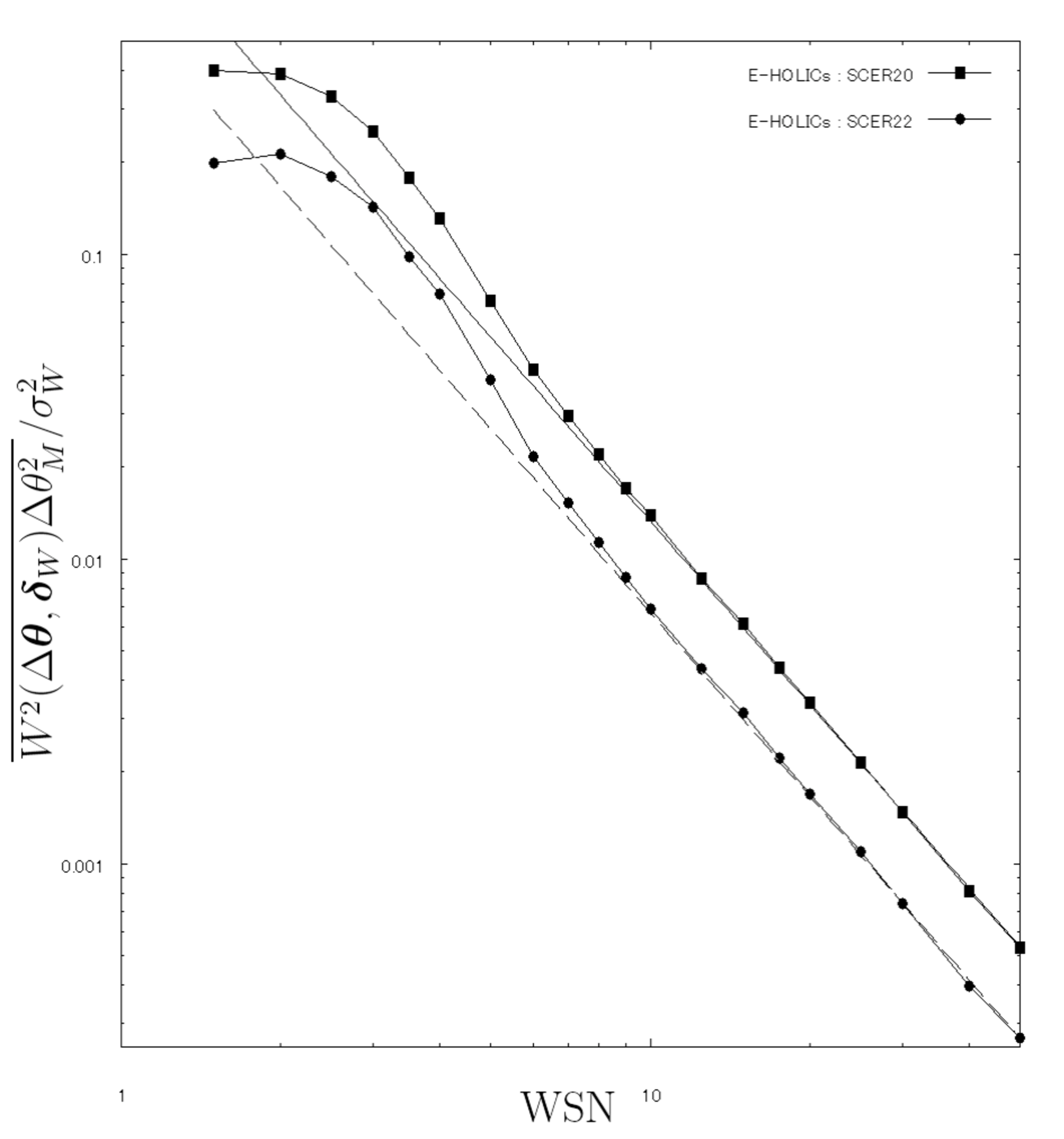}
\caption{
\label{fig:E-HOLICs_3_OEW_C}
Plots are $\overline{\Delta\theta^2_M}/\sigma_W^2$ with elliptical Gaussian image $\bde_I=(0.5, 0)$ and E-HOLICs method with observed ellipticity for weight function.
Horizontal axis means WSN and vertical axis means $\overline{\Delta\theta^2_M}/\sigma_W^2$.
Square(circle) means $\overline{\Delta\theta^2_0}/\sigma_W^2(\overline{\Delta\theta^2_2}/\sigma_W^2)$ measured from simulation data and solid(dashed) line means analytical prediction i.e. eq.(\ref{eq:DG2DT20GaussEH})(eq.(\ref{eq:DG2DT22GaussEH})).
} 
\end{figure*}

If the object is EGI, the averaged CSEs are calculated as follows.
\begin{eqnarray}
\label{eq:DG2DT20GaussEH}
\overline{\Delta\theta^2_0}\approx\overline{ W^2(\Delta\bth,\bde_W)\Delta\theta^2_0}&\approx&\frac{1}{WSN^2}\frac{\sigma_W^2}{1-\delta_I^2}\\
\label{eq:DG2DT22GaussEH}
\overline{\Delta\theta^2_2}\approx\overline{ W^2(\Delta\bth,\bde_W)\Delta\theta^2_2}&\approx&\frac{1}{WSN^2}\frac{\sigma_W^2}{1-\delta_I^2}\bde_I\\
\bde_C=\bde_I.
\end{eqnarray}
Fig.\ref{fig:E-HOLICs_3_OEW_C} shows the comparison between numerical results and theoretical prediction given by eq.(\ref{eq:DG2DT20GaussEH}) and eq.(\ref{eq:DG2DT22GaussEH}) with $\bde_I=(0.5, 0)$.
\subsubsection{Complex Moments}
In this section, we calculate the effects of $\Delta\bde$ for the complex moments.

$W\lr{\bth,\bde_W}$ can be expanded up to 2nd order in RCN as
\begin{eqnarray}
W\lr{\bth,\bde_W}&=&W\lr{\bth,\bde_I+\Delta\bde_{(1)}+\Delta\bde_{(2)}}
\nonumber\\&\approx&
W(\bth,\bde_I)\lr{1+\frac{\Real{\lr{\Delta\bde_{(1)}+\Delta\bde_{(2)}}^*\theta^2_2}}{\sigma_W^2}+\frac{\Real{\Delta\bde_{(1)}^*\theta^2_2}^2}{2\sigma_W^4}}
\end{eqnarray}
By using this expanded weight, $\hat Z^N_M(I^{obs},\bde_W)$ can be expressed by $\Delta\bde$ as
\begin{eqnarray}
\label{eq:hatZNMDd}
\hat Z^N_M(I^{obs},\bde_W)&\approx&Z^N_0(I^{obj},\bde_I)\Bigglrs{
\cH^N_M+\Delta\cH^N_{M(1)}(\bde_I)+\Delta\cH^N_{M(2)}(\bde_I)
\nonumber\\&&
+\frac{1}{2\sigma_W^2}\lr{\lr{\Delta\bde_{(1)}+\Delta\bde_{(2)}}^*\cH^{N+2}_{M+2}+\lr{\Delta\bde_{(1)}(\bde_I)+\Delta\bde_{(2)}(\bde_I)}\cH^{N+2}_{M-2}}
\nonumber\\&&
+\frac{1}{2\sigma_W^2}\lr{\Delta\bde_{(1)}^*\Delta\cH^{N+2}_{M+2(1)}+\Delta\bde_{(1)}\Delta\cH^{N+2}_{M-2(1)}}
\nonumber\\&&
+\frac{1}{8\sigma_W^4}\lr{\Delta\bde_{(1)}^{*2}\cH^{N+4}_{M+4}+2|\Delta\bde_{(1)}|^2\cH^{N+4}_{M}+\Delta\bde_{(1)}^2\cH^{N+4}_{M-4}}
}(I^{obj},Z^N_0,\bde_I).
\end{eqnarray}
We postpone to calculate eq.(\ref{eq:hatZNMDd}) in detail until an appropriate expression for $\Delta\bde$ 
is available later.

\subsubsection{Ellipticity Without PSF Correction}
In this section we calculate the systematic error in measuring the ellipticity,
but we don't consider PSF smearing. 
This situation is achieved in space observation such as HST, and thus useful.
Systematic error in measuring ellipticity of objects with PSF smearing can be obtained with more complex calculations and we make a comment in the next section.

Without PSF smearing, we use the observed ellipticity for the weight function (i.e. $\bde_W=\hat\cH^2_2(I^{obs},Z^2_0,\bde_W)$).
By expanding $\hat\cH^2_2$ by $\Delta\bde$, we obtain
\begin{eqnarray}
\bde_W&=&\hat\cH^2_2(I^{obs},Z^2_0,\bde_W)\approx
\cH^2_2(I^{obj},Z^2_0,\bde_I)+\Delta {Z_R}^2_2
-\frac14\lr{\Delta\bde_{(1)}\cC0^2_{0-}+\Delta\bde^*_{(1)}\cC0^2_{4-}}
\nonumber\\&&
+\Delta\cH^2_{2(2)} - \bde_I\Delta\cH^2_{0(2)}
-\frac14\lr{\Delta\bde_{(2)}\cC0^2_{0-}+\Delta\bde^*_{(2)}\cC0^2_{4-}}
\nonumber\\&&
+\frac{1}{2\sigma_W^2}\lr{\Delta\bde_{(1)}\Delta {Z_R}^4_0+\Delta\bde^*_{(1)}\Delta {Z_R}^4_4}
\nonumber\\&&
-\frac{\cH^4_0(\Iobj,Z^2_0,\bde_I)}{16\sigma_W^4}\lr{\Delta\bde^2_{(1)}\cC0^4_{-2-}+2|\Delta\bde_{(1)}|^2\cC0^4_{2-}+\Delta\bde^{*2}_{I(1)}\cC0^4_{6-}}
\nonumber\\&&
-\lr{\Delta {Z_R}^2_2-\frac14\lr{\Delta\bde_{(1)}\cC0^2_{0-}+\Delta\bde^*_{(1)}\cC0^2_{4-}}}
\nonumber\\&&\times
\lr{\frac{Z^2_0(\IRCN,\bde_I)}{Z^2_0(\Iobj,\bde_I)}+\frac{1}{2\sigma_W^2}\lr{\Delta\bde_{(1)}\cH^{4*}_2(\Iobj,Z^2_0,\bde_I)+\Delta\bde^*_{(1)}\cH^4_2(\Iobj,Z^2_0,\bde_I)}},
\end{eqnarray}
where
\begin{eqnarray}
\Delta {Z_R}^N_M \equiv \frac{Z^N_M(\IRCN,\bde_I)}{Z^2_0(\Iobj,\bde_I)} - \bde_I \frac{Z^N_{M-2}(\IRCN,\bde_I)}{Z^2_0(\Iobj,\bde_I)}.
\end{eqnarray}
$\Delta\bde_{(X)}$ are obtained by comparing 1st order as 
\begin{eqnarray}
\frac14\lr{\cC4^2_{0-}\Delta\bde_{(1)}+\cC0^2_{4-}\Delta\bde^*_{(1)}}&=&\DZR^2_2
\nonumber\\
\label{eq:Bbde1EH}
\Delta\bde_{(1)}&=&
4\lr{\tcC4^{2*}_{0-}\DZR^2_2-\tcC0^2_{4-}\DZR^{2*}_2}
\end{eqnarray}
where
\begin{eqnarray}
\tcC4^2_{0-}&\equiv&\frac{\cC4^2_{0-}}{|\cC4^2_{0-}|^2-|\cC0^2_{4-}|^2}
\nonumber\\
\tcC0^2_{4-}&\equiv&\frac{\cC0^2_{4-}}{|\cC4^2_{0-}|^2-|\cC0^2_{4-}|^2}
\end{eqnarray}
and 2nd order as
\begin{eqnarray}
\label{eq:Bbde2EH}
\Delta\bde_{(2)}&=&
\Bigglrs{\Delta\cH^2_{2(2)}(\bde_I)-\bde_I\Delta\cH^2_{0(2)}(\bde_I)}
-\frac{1}{4}\lr{\Delta\bde_{(2)}\cC0^2_{0-}+\Delta\bde^*_{(2)}\cC0^2_{4-}}
\nonumber\\&&
-\frac{\cH^{4}_{0}(I^{obj},Z^2_0,\bde_I)}{16\sigma_W^2}\lr{\Delta\bde_{(1)}^{2*}\cC0^4_{6-}
+2|\Delta\bde_{(1)}|^2 \cC0^4_{2-}
+\Delta\bde_{(1)}^2   \cC0^{4}_{-2-}}
\nonumber\\&&
+\frac{1}{2\sigma_W^2}\lr{\Delta\bde_{(1)}  \DZR^4_0+\Delta\bde_{(1)}^*\DZR^4_4}
-\Delta\bde_{(1)}\lr{\frac{Z^2_0(I^{RCN},\bde_I)}{Z^2_0(I^{obj},\bde_I)}+\frac{\Real{\Delta\bde_{(1)}^*\cH^{4}_{2}(I^{obj},Z^2_0,\bde_I)}}{\sigma_W^2}}
\end{eqnarray}
and the average is given by 
\begin{eqnarray}
\label{eq:ABbde2EH}
\overline{\frac14\lr{\cC4^2_{0-}\Delta\bde_{(2)}+\cC0^2_{4-}\Delta\bde^*_{(2)}}}&\approx&
\Biglr{\Delta\cH^2_{2(2)}(\bde_I)-\bde_I\Delta\cH^2_{0(2)}(\bde_I)}
\nonumber\\&&
+\frac{\lr{\sigma_W^2\cH^0_0(I^{obj},Z^2_0,\bde_I)}^2}{4\lr{1-\delta_I^2}}\frac{1}{\WSN^2}\Biglr{\bde_I\tcC4^2_{0-}+3\bde_I^*\tcC0^2_{4-}}
\nonumber\\&&\hspace{-150pt}
-\frac{\lrs{\cH^4_0\lr{\sigma_W^2\cH^0_0}^2}(I^{obj},Z^2_0,\bde_I)}{8\sigma_W^2\lr{1-\delta_I^2}}\frac{1}{\WSN^2}\Bigglr{
2\cC4^4_{2-}\lr{2\lr{|\tcC4^2_{0-}|^2+|\tcC0^2_{4-}|^2}-\lrabs{\tcC4^2_{0-}\bde_I-\tcC0^2_{4-}\bde^*}^2}
\nonumber\\&&\hspace{-150pt}
-\cC8^4_{-2-}\lr{4\tcC4^2_{0-}\tcC0^2_{4-}   +\lr{\tcC4^2_{0-}\bde_I-\tcC0^2_{4-}\bde^*}^2}
-\cC0^4_{  6-}\lr{4\tcC4^2_{0-}\tcC0^{2*}_{4-}+\lr{\tcC4^2_{0-}\bde_I-\tcC0^2_{4-}\bde^*}^{2*}}}
\end{eqnarray}
In deriving this expression we used an approximation on the relation of the phase angle(e.g. $\bde^*\cH^4_2 \approx \bde\cH^{4*}_2$ see Okura and Futamase 2012).

\begin{figure*}[htbp]	
\epsscale{1.0}
\plotone{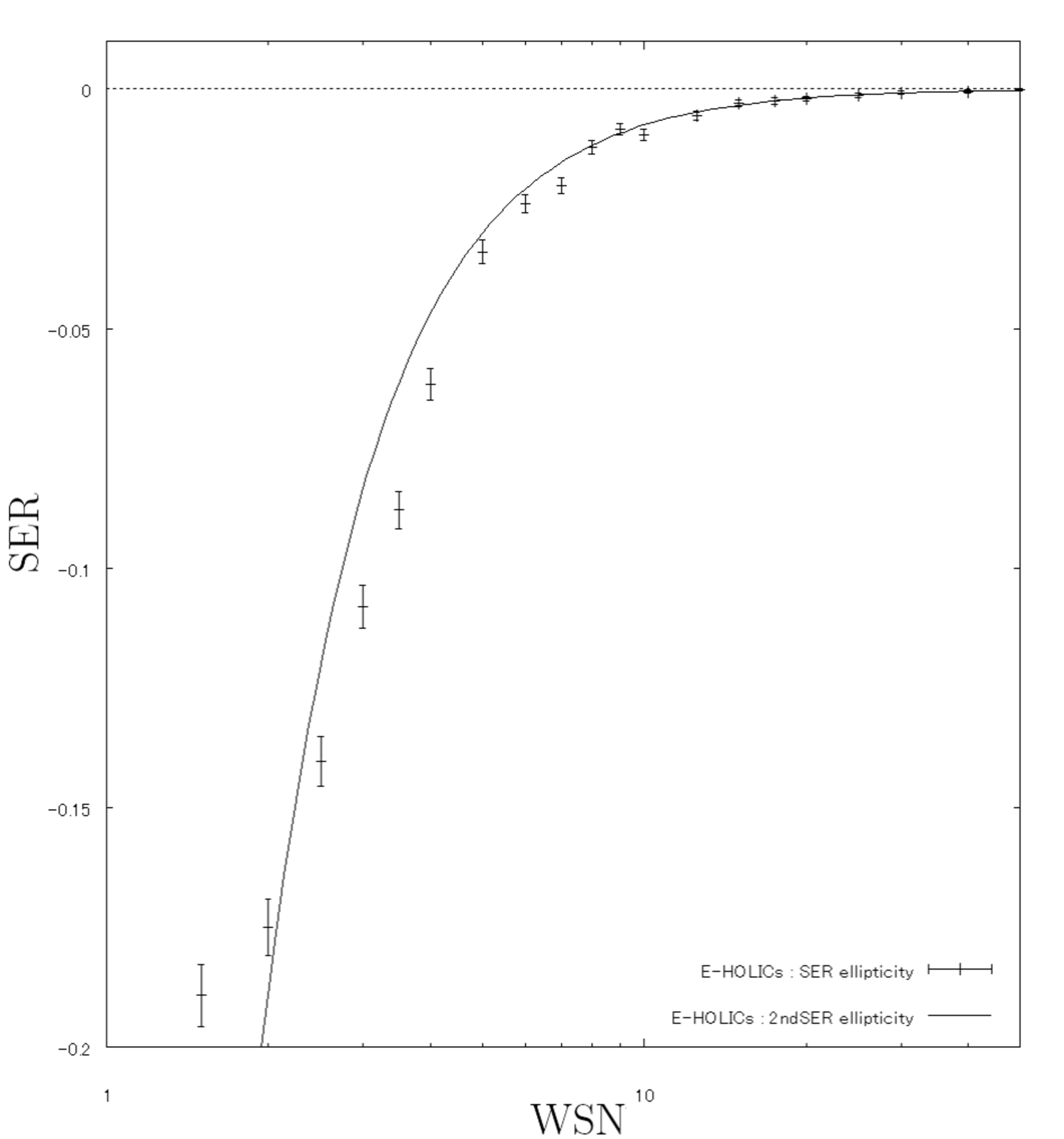}
\caption{
\label{fig:E-HOLICs_3_OEW_EReal}
SER due to RCN for observed ellipticity in the situation of EGI and E-HOLICs.
Horizontal axis means WSN and vertical axis means systematic error $\overline{\Delta \bde_{(2)}}/\bde_I$.
Plots are simulated results and line is estimation by eq.(\ref{eq:OEW_DD2}).
} 
\end{figure*}

In the situation of EGI, eq.(\ref{eq:ABbde2EH}) is reduced to the following expression.
\begin{eqnarray}
\label{eq:REALEGOEW}
\overline{\bde_W}&=&\lr{1-\frac{2\lr{1-\delta_I^2}}{WSN^2}}\bde_I\\
\label{eq:OEW_DD2}
\overline{\Delta\bde_{(2)}}&=&-\frac{2\lr{1-\delta_I^2}}{\WSN^2}\bde_I
\end{eqnarray}
Figure \ref{fig:E-HOLICs_3_OEW_EReal} shows the comparison between numerical result of SER and theoretical prediction 2ndSER given by eq.(\ref{eq:OEW_DD2}).

By using $\overline{\Delta\bde_{(2)}}$ and eq.(\ref{eq:hatZNMDd}), the average of $\Delta\cH^N_{M(2)}(\bde_W)$ is obtained as
\begin{eqnarray}
\label{eq:EH_AVEDHNM}
\overline{\Delta\cH^N_{M(2)}(\bde_W)}
&\approx&\Bigglrs{\overline{\Delta\cH^N_{M(2)}(\bde_I)}
+\frac{1}{2\sigma_W^2}\lr{\overline{\Delta\bde_{(2)}}^*\cH^{N+2}_{M+2}+\overline{\Delta\bde_{(2)}(\bde_I)}\cH^{N+2}_{M-2}}
\nonumber\\&&
+\frac{1}{2\sigma_W^2}\lr{\overline{\Delta\bde_{(1)}^*\Delta\cH^{N+2}_{M+2(1)}}+\overline{\Delta\bde_{(1)}\Delta\cH^{N+2}_{M-2(1)}}}
\nonumber\\&&
+\frac{1}{8\sigma_W^2}\lr{\overline{\Delta\bde_{(1)}^{*2}}\cH^{N+4}_{M+4}+2\overline{|\Delta\bde_{(1)}|^2}\cH^{N+4}_{M}+\overline{\Delta\bde_{(1)}^2}\cH^{N+4}_{M-4}}
}(I^{obj},Z^N_0,\bde_I)
\nonumber\\&=&
\Bigglrs{
\overline{\Delta\cH^N_{M(2)}(\bde_I)}
+\frac{1}{2\sigma_W^2}\lr{\overline{\Delta\bde_{(2)}}^*\cH^{N+2}_{M+2}+\overline{\Delta\bde_{(2)}(\bde_I)}\cH^{N+2}_{M-2}}
\nonumber\\&&
\frac{\cH^0_0(\Iobj,Z^2_0,\bde_I)}{\WSN^2}\frac{2\cH^0_0}{\sigma_W^2S_W}\Biglr{\tcC4^2_{0-}\lr{\overline{G^{N+4}_M}-\bde_I^*\overline{G^{N+4}_{M+2}}}+\tcC0^{2*}_{4-}\lr{\overline{G^{N+4}_{M+4}}-\bde_I\overline{G^{N+4}_{M+2}}}
\nonumber\\&&\hspace{125pt}
+\tcC4^{2*}_{0-}\lr{\overline{G^{N+4}_M}-\bde_I\overline{G^{N+4}_{M-2}}}+\tcC0^{2}_{4-}\lr{\overline{G^{N+4}_{M-4}}-\bde^*_I\overline{G^{N+4}_{M-2}}}
}
\nonumber\\&&\hspace{-50pt}
+\frac{\lr{\cH^0_0(\Iobj,Z^2_0,\bde_I)}^2}{4\WSN^2\lr{1-\delta_I^2}}\Bigglr{
2\cH^{N+4}_M\lr{\lr{|\tcC4^{2*}_{0-}|^2+|\tcC0^2_{4-}|^2}\lr{2-\delta_I^2}+\tcC4^{2*}_{0-}\tcC0^{2*}_{4-}\bde_I^2+\tcC4^{2}_{0-}\tcC0^{2}_{4-}\bde_I^{2*}}
\nonumber\\&&\hspace{25pt}
-\cH^{N+4}_{M+4}\lr{2\cC4^2_{0-}\tcC0^{2*}_{4-}\lr{2-\delta_I^2}+\lr{\tcC4^2_{0-}}^2\bde_I^2+\lr{\tcC0^{2*}_{4-}}^2\bde_I^{2*}}
\nonumber\\&&\hspace{25pt}
-\cH^{N+4}_{M-4}\lr{2\cC4^{2*}_{0-}\tcC0^2_{4-}\lr{2-\delta_I^2}+\lr{\tcC4^{2*}_{0-}}^2\bde_I^{2*}+\lr{\tcC0^2_{4-}}^2\bde_I^2}}
}(I^{obj},Z^N_0,\bde_I)
\end{eqnarray}

\begin{figure*}[htbp]	
\epsscale{1.0}
\plotone{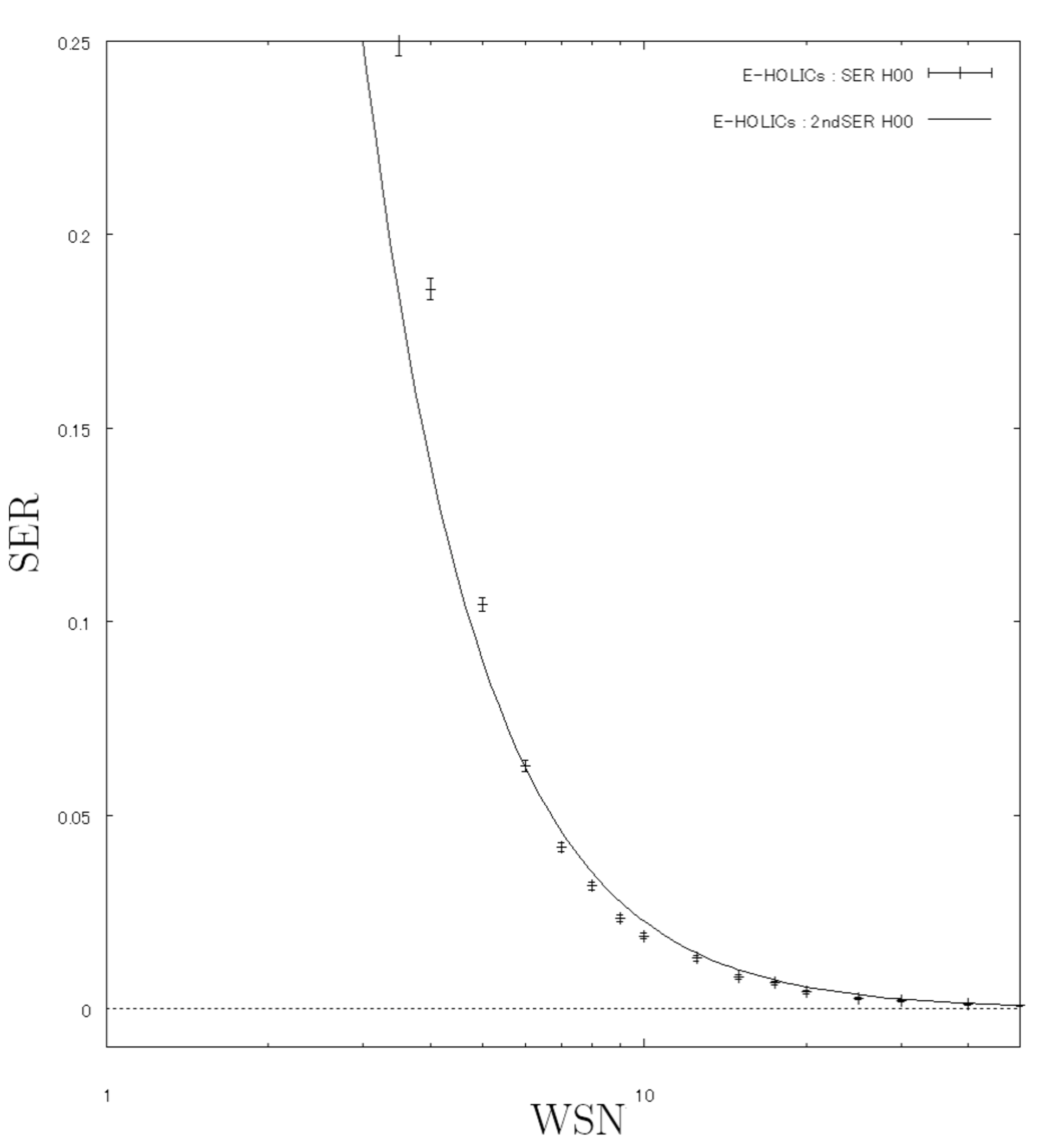}
\caption{
\label{fig:E-HOLICs_3_OEW_H0}
SER due to RCN for monopole moments measured with EGI and E-HOLICs method.
Horizontal axis means WSN and vertical axis means systematic error.
Plots are simulated result of SER of $\Delta\cH^0_0$ and line is estimated value from eq.(\ref{eq:EH_AVEDHNM}).
} 
\end{figure*}

\begin{figure*}[htbp]	
\epsscale{1.0}
\plotone{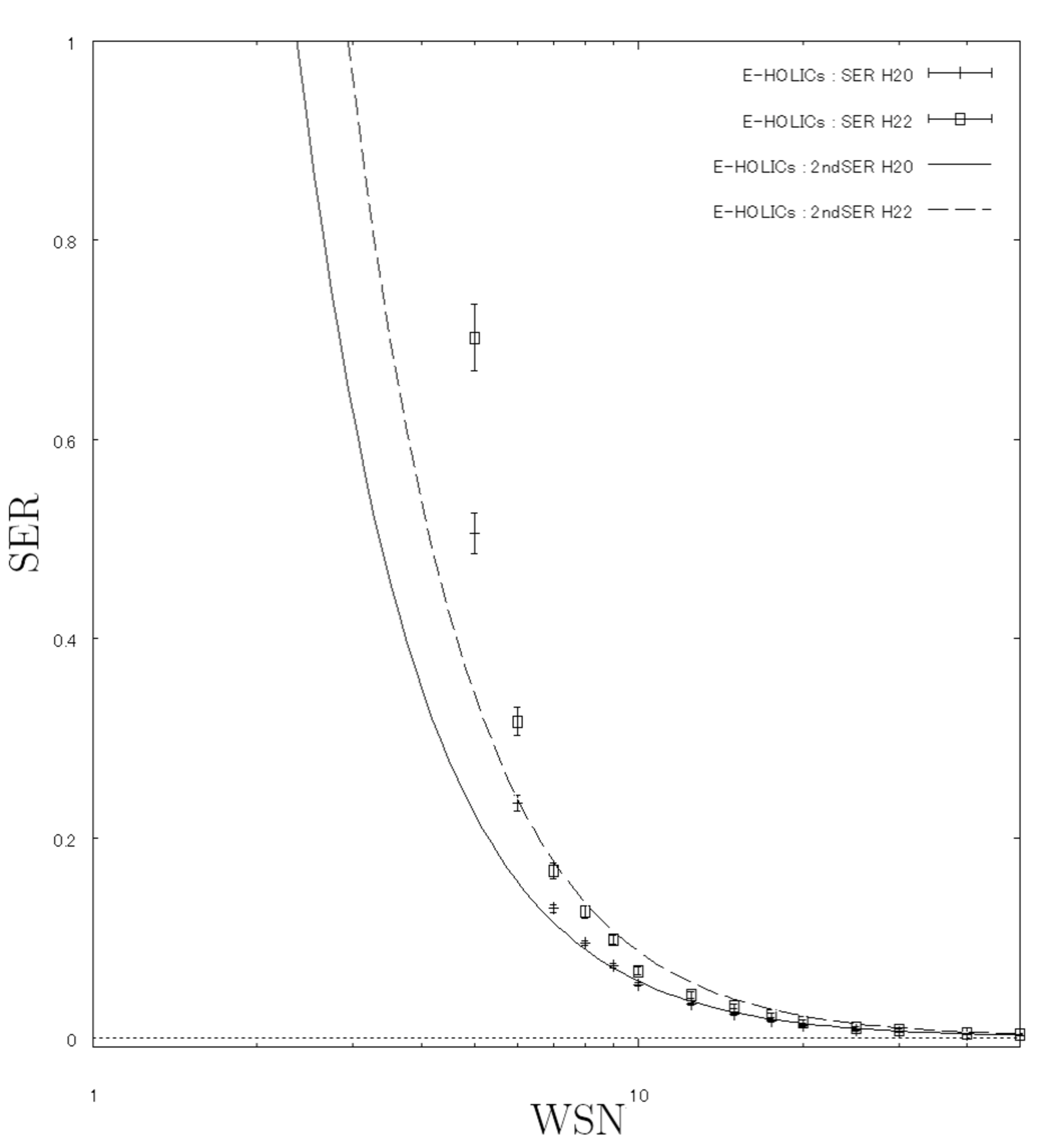}
\caption{
\label{fig:E-HOLICs_3_OEW_H2}
SER due to RCN for quadrupole moments measured with EGI and E-HOLICs method.
Horizontal axis means WSN and vertical axis means systematic error.
Cross(square) plots are simulated results of SER $\Delta\cH^2_{0(2)}$($\Delta\cH^2_{2(2)}$) and  solid(long dash) line is estimated value from eq.(\ref{eq:EH_AVEDHNM})} 
\end{figure*}

\begin{figure*}[htbp]	
\epsscale{1.0}
\plotone{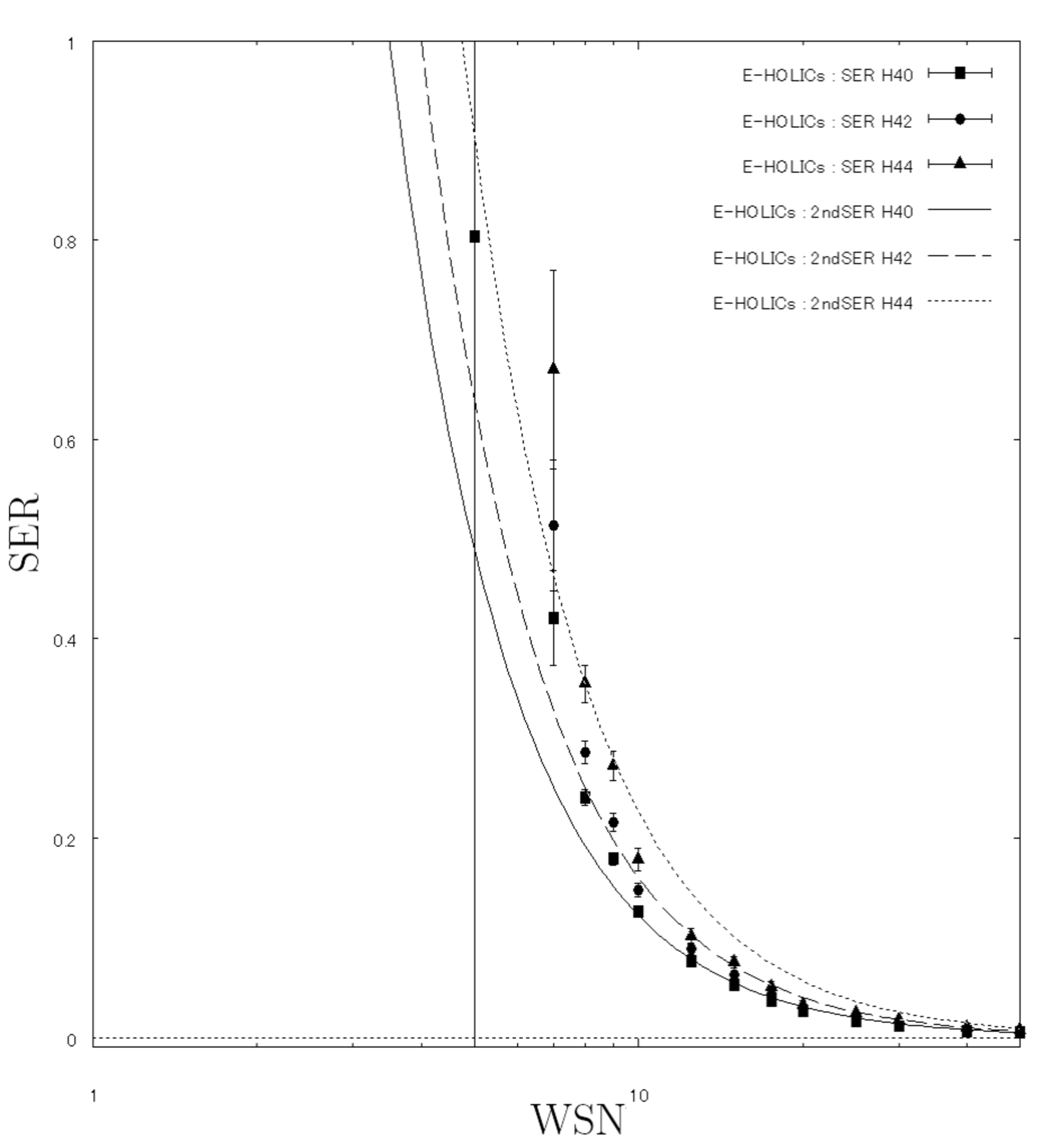}
\caption{
\label{fig:E-HOLICs_3_OEW_H4}
SER due to RCN for quadrupole moments measured with EGI and E-HOLICs method.
Horizontal axis means WSN and vertical axis means systematic error.
Square(circle, triangle) plots are simulated results of SER of $\Delta\cH^4_{0(2)}$($\Delta\cH^4_{2(2)}$, $\Delta\cH^4_{4(2)}$) and solid(long dash, dash) line is estimated value from eq.(\ref{eq:EH_AVEDHNM}).
} 
\end{figure*}

Fig.\ref{fig:E-HOLICs_3_OEW_H0} to Fig.\ref{fig:E-HOLICs_3_OEW_H4} are simulated results of SER of Complex moments. 
We can see that the systematic error of complex moments can be almost estimated by eq.(\ref{eq:EH_AVEDHNM}) expect for the objects with low WSN,
but there are differences between simulation and theory in the estimation of high spin moments $\cH^4_4$.

\subsubsection{Tests using GREAT08 simulation image}
We test the correction formulas obtained in the previous section using  GREAT08 simulation data.
First, we selected 2 objects from "LowNoise\_Known set0001.fits" (we call Sample A and Sample B),
and we compared complex moments and ellipticity between the original and noisy objects, where noisy objects are created by adding 10000 types(having same RMS) of RCN to the original object.
Table \ref{tab:GREAT08C_SampleA} and Table \ref{tab:GREAT08C_SampleB} are the results of the tests.
The 2nd column of the table shows 
the average of normalized differences (i.e. the ratio of systematic error with and without the corrections)
SER and 1$\sigma$ error
and the 3rd column shows corrected SER which means systematic error after corrected by 2ndSER
 and 1$\sigma$ error
From these tables, we can see the averaged errors of ellipticity with correction are under 1\%.
\begin{table}[htbp]
\begin{center}
\begin{tabular}{|c|c|c|} \hline
Parameter  & SER & corrected SER \\ \hline
$Z^0_0$          &   0.0386 $\pm$0.00104  &   0.00664 $\pm$0.00119 \\ \hline
$Z^2_0$          &     0.171 $\pm$0.00426  &     0.0118 $\pm$0.00558 \\ \hline
$\Real{Z^2_2}$ &    0.233 $\pm$0.00841  &     0.0253 $\pm$0.00911 \\ \hline
$\Img{Z^2_2}$  &     0.237 $\pm$0.00832 &     0.0189 $\pm$0.0109 \\ \hline
$Z^4_0$          &     0.491 $\pm$0.0133   &   -0.0625 $\pm$0.0235 \\ \hline
$\Real{Z^4_2}$ &    0.561 $\pm$0.0172    &   -0.0632 $\pm$0.0233 \\ \hline
$\Img{Z^4_2}$  &     0.583 $\pm$0.0180   &   -0.0876 $\pm$0.0363 \\ \hline
$\Real{Z^4_4}$ &      1.24 $\pm$0.224     &     -0.286 $\pm$0.556 \\ \hline
$\Img{Z^4_4}$  &     0.707 $\pm$0.0235   &   -0.0682 $\pm$0.0387 \\ \hline
$\Real{\bde}$  & -0.0225 $\pm$0.00402  & -0.00636 $\pm$0.00428 \\ \hline
$\Img{\bde}$   & -0.0199 $\pm$0.00373  & -0.00661 $\pm$0.00400 \\ \hline
\end{tabular} 
\caption{
\label{tab:GREAT08C_SampleA}
SER and corrected SER of sample A, average of WSN of noised object is 9.1 and ellipticity is (0.39,-0.42)} 
\end{center}
\end{table} 
\begin{table}[htbp]
\begin{center}
\begin{tabular}{|c|c|c|} \hline
Parameter  & SER & corrected SER \\ \hline
$Z^0_0$          &  0.0380 $\pm$0.00110  &  0.00593 $\pm$0.00118 \\ \hline
$Z^2_0$          &    0.178 $\pm$0.00469  &    0.0104 $\pm$0.00473 \\ \hline
$\Real{Z^2_2}$ &   0.231 $\pm$0.00740  &    0.0130 $\pm$0.00718 \\ \hline
$\Img{Z^2_2}$  &    0.225 $\pm$0.0105   &   0.00531 $\pm$0.00985 \\ \hline
$Z^4_0$          &    0.495 $\pm$0.0133   &   -0.0561 $\pm$0.0130 \\ \hline
$\Real{Z^4_2}$ &   0.551 $\pm$0.0157    &   -0.0824 $\pm$0.0152 \\ \hline
$\Img{Z^4_2}$  &    0.539 $\pm$0.0191   &     -0.107 $\pm$0.0181 \\ \hline
$\Real{Z^4_4}$ &    0.648 $\pm$0.0221   &     -0.105 $\pm$0.0207 \\ \hline
$\Img{Z^4_4}$  &    0.628 $\pm$0.0229   &     -0.142 $\pm$0.0191 \\ \hline
$\Real{\bde}$  & -0.0254 $\pm$0.00239  & -0.00256 $\pm$0.00252 \\ \hline
$\Img{\bde}$   & -0.0197 $\pm$0.00650  & -0.00917 $\pm$0.00676 \\ \hline
\end{tabular} 
\caption{
\label{tab:GREAT08C_SampleB}
SER and corrected SER of sample B, average of WSN of noised object is 9.1 and ellipticity (0.60, -0.23).}
\end{center}
\end{table} 

\section{Conclusion and Future works}
Following the previous work we studied in this paper the systematic
error caused by signal to noise(SN) ratio  
of the observed image in our weak lensing analysis called E-HOLICs. 
It has been known that the shear is underestimated when low SN
background images are used and is overestimated when high SN background
images are used in the weak lensimng analysis.  
The latter error was improved in the previous work. The improvement of the former error is important because 
if we have such improvement, we will be able to use many faint
background sources which improves the statistical accuracy of the weak
lensing analysis.  

We have identified the origin of the systematic error as the photon random count noise by sky. 
Although its 1st order effect vanishes by averaging, but 2nd order effects are not canceled in measuring 
the moments and centroid of the images. We investigated this effect 
carefully and obtain the formulas in KSB method and E-HOLICs method to
correct the effect in measuring moments and ellipticity. 
Although general expressions for these formulas are complicated, they
reduce to relatively simple forms for images with an elliptical Gaussian form(EGI). 
We tested the validity of the correction formula eq.(\ref{eq:cformH}) and eq.(\ref{eq:cformD}) for EGIs using simulatoin data.
Furthermore we applied the general formulas to GREAT08 and confirmed that the systematic error reduces to less than 1\% in measuring ellipticity for images with WSN=9.1 which roughly corresponds to SN=3 object. 

Although the present analysis has not taken into account the Point
Spread function(PSF) correction which is necessary for the observation
from the ground. PSF correction uses complicated combinations of higher
moments and will be very complicated in E-HOLICs approach. However the above result is very encouraging and is worthwhile challenging.
Finally we should point out that the present work will be applicable for
the space based observation because PSF by instrument is expected to be
small for such observation. It will be very interesting to confirm this expectation by
using data such as COSMOS. 

We thank Satoshi Miyazaki, Takashi hamana, Keiichi Umetsu,  Nobuhiro
Okabe, Yousuke Utsumi and Yuichi Higuchi for useful discussions and comments. 
This work is partially supported by the COE program''Weaving Science Web
beyond Particle-matter Hierarchy'' at Tohoku University and  Grant-in
-Aid for Scientific Research from JSPS(Nos.18072001, 23540282 for TF). 
\appendix
\section{Appendix}
\subsection{$\overline{G^N_M(\bde_W)}$}
\label{AP:GNM}
Integration of square Gaussian
\begin{eqnarray}
\overline{G^0_0(\bde_W)}&=& \frac12\frac{\pi\sigma_W^2}{\sqrt{1-\delta_W^2}}=\frac{S_W}{2}\\
\overline{G^2_0(\bde_W)}&=& \frac12\frac{\sigma^2_W}{1-\delta_W^2}\frac{S_W}{2}=\frac12\frac{\sigma^2_W}{1-\delta_W^2}\overline{G^0_0}\\
\overline{G^2_2(\bde_W)}&=& \frac12\bde_W\frac{\sigma^2_W}{1-\delta_W^2}\frac{S_W}{2}=\frac12\frac{\sigma^2_W}{1-\delta_W^2}\overline{G^0_0}\bde_W=\bde_W \overline{G^2_0}\\
\overline{G^4_0(\bde_W)}&=& \frac14\frac{\lr{2+\delta_W^2}\sigma^4_W}{\lr{1-\delta_W^2}^2}\frac{S_W}{2}=\frac12\frac{(2+\delta_W^2)\sigma_W^2}{1-\delta_W^2}\overline{G^2_0}\\
\overline{G^4_2(\bde_W)}&=& \frac14\frac{3\bde_W\sigma^4_W}{\lr{1-\delta_W^2}^2}\frac{S_W}{2}=\frac12\frac{3\bde_W\sigma_W^2}{1-\delta_W^2}\overline{G^2_0}\\
\overline{G^4_4(\bde_W)}&=& \frac14\frac{3\bde_W^2\sigma^4_W}{\lr{1-\delta_W^2}^2}\frac{S_W}{2}\\
\overline{G^6_0(\bde_W)}&=&\frac38\frac{\lr{2+3\delta_W^2}\sigma^6_W}{\lr{1-\delta_W^2}^3}\frac{S_W}{2}\\
\overline{G^6_2(\bde_W)}&=&\frac38\frac{\lr{4+ \delta_W^2}\bde_W\sigma^6_W}{\lr{1-\delta_W^2}^3}\frac{S_W}{2}\\
\overline{G^6_4(\bde_W)}&=&\frac38\frac{5\bde^2_W\sigma^6_W}{\lr{1-\delta_W^2}^3}\frac{S_W}{2}\\
\overline{G^6_6(\bde_W)}&=&\frac38\frac{5\bde^3_W\sigma^6_W}{\lr{1-\delta_W^2}^3}\frac{S_W}{2}\\
\overline{G^8_0(\bde_W)}&=&\frac{3}{16}\frac{\lr{8+24\delta_W^2+3\delta_W^4}\sigma^8_W}{\lr{1-\delta_W^2}^4}\frac{S_W}{2}\\
\overline{G^8_2(\bde_W)}&=&\frac{3}{16}\frac{5\lr{4+3\delta_W^2}\bde_W\sigma^8_W}{\lr{1-\delta_W^2}^4}\frac{S_W}{2}\\
\overline{G^8_4(\bde_W)}&=&\frac{3}{16}\frac{5\lr{6+\delta_W^2}\bde_W^2\sigma^8_W}{\lr{1-\delta_W^2}^4}\frac{S_W}{2}\\
\overline{G^8_6(\bde_W)}&=&\frac{3}{16}\frac{15\bde_W^3\sigma^8_W}{\lr{1-\delta_W^2}^4}\frac{S_W}{2}
\\
\overline{G^8_8(\bde_W)}&=&\frac{3}{16}\frac{15\bde_W^4\sigma^8_W}{\lr{1-\delta_W^2}^4}\frac{S_W}{2}
\end{eqnarray}
\subsection{C Coefficients in KSB method with Elliptical Gaussian Image}
\label{AP:CNM_KSB}
In the situation of EGI and KSB method, C coefficients are obtained analytically as 
\begin{eqnarray}
\cC^N_{N\pm} &=& -2\cH^N_N(\Iobj,Z^N_0,0)
\nonumber\\\hspace{-0pt}
\cC^N_{0\pm} &=& -2\delta^2_{I,0} \hspace{50pt} (N\neq0)
\nonumber\\\hspace{-0pt}
\cC^{4}_{2+}&=&-2\delta_{I,0}^2\cH^4_2(I^{obj},Z^4_0,0)
\nonumber\\\hspace{-0pt}
\cC^{4}_{2-}&=&-\frac23\lr{2+\delta_{I,0}^2}\cH^4_2(I^{obj},Z^4_0,0)
\nonumber\\\hspace{-0pt}
\cCX^N_{M\pm} &=& \lr{X-\frac{2+N+M}{2}-\lr{1+N}\delta_{I,0}^2}\cH^N_M(\Iobj,Z^N_0,0)
\nonumber\\\hspace{-0pt}
\cC0^N_{N+2\pm} &=& -\frac2{\sigma_W^2}\cH^{N+2}_{N+2}(I^{obj},Z^N_0,0)
\end{eqnarray}
but following equation cannot be adopt upper general forms
\begin{eqnarray}
\cCX^4_{0\pm} &=& \lr{X-3-6\delta_{I,0}^2}\cH^N_M(\Iobj,Z^N_0,0)
\end{eqnarray}
where $N = 0, 2$ or $4$ and $|M| \leq |N|$.

\subsection{C Coefficients in E-HOLICs method with true Ellipticity for Weight function and Elliptical Gaussian Image}
\label{AP:CNM_TEH}
In the situation of EGI and E-HOLICs method with true ellipticity, C coefficients are obtained analytically as 
\begin{eqnarray}
\cC^N_{N\pm} &=& -\lr{1\pm1}\lr{1-\delta_I^2}\cH^N_N(\Iobj,Z^N_0,\bde_I)
\nonumber\\\hspace{-0pt}
\cC^{N}_{M\pm}&=&0  \hspace{50pt} (N\neq M)
\nonumber\\\hspace{-0pt}
\cCX^{N}_{M\pm}&=&\lr{X-\frac{2+N\pm M}{2}}\cH^N_M(I^{obj},Z^N_0,\bde_I)
\nonumber\\\hspace{-0pt}
\cC0^N_{N+2\pm} &=& -\frac{1\pm1}{\sigma_W^2}\cH^{N+2}_{N+2}(I^{obj},Z^N_0,0)
\end{eqnarray}
but following equation cannot be adopt upper general forms
\begin{eqnarray}
\cC^0_{0-} &=& -2\lr{1-\delta_I^2}\cH^0_0(\Iobj,Z^0_0,\bde_I)
\end{eqnarray}
where $N = 0, 2$ or $4$ and $|M| \leq |N|$.
\begin{eqnarray}
\tcC2^0_{0\pm} &=& \frac{\cC2^0_{0\pm}}{|\cC2^0_{0+}|^2-|\cC0^0_{2-}|^2}=1\\
\tcC0^0_{2-    } &=& \frac{\cC0^0_{2-    }}{|\cC2^0_{0+}|^2-|\cC0^0_{2-}|^2}=0\\
\tcC4^2_{0-}&=&\frac{\cC4^2_{0-}}{|\cC4^2_{0-}|^2-|\cC0^2_{4-}|^2}=\frac12\\
\tcC0^2_{4-}&=&\frac{\cC0^2_{4-}}{|\cC4^2_{0-}|^2-|\cC0^2_{4-}|^2}=0
\end{eqnarray}

\end{document}